\documentclass[twocolumn,10pt]{IEEEtran}

\usepackage{amsmath}
\usepackage{amsfonts}
\usepackage{mathrsfs}
\usepackage{dsfont}
\usepackage{eucal}
\usepackage{pifont}
\usepackage{float}
\usepackage{amssymb}
\usepackage{verbatim}
\usepackage{upgreek}
\usepackage{color}
\usepackage[final]{graphicx}
\usepackage{subfigure}
\usepackage{epsfig}
\usepackage{booktabs}
\usepackage{bm}
\usepackage{setspace}
\usepackage[latin1]{inputenc}
\usepackage{algorithm}
\usepackage{float} 
\usepackage{colortbl}
\usepackage{xcolor} 
\usepackage{multirow}
\usepackage{rotating}
\usepackage{cite}
\usepackage{balance}
\usepackage[colorlinks]{hyperref}

\begin{document}

\title{Monitoring of Critical Infrastructures \\by Micro-Motion Estimation:\\ the Mosul Dam Destabilization}

\author{Filippo Biondi, \IEEEmembership{Member, IEEE}, Pia Addabbo, \IEEEmembership{Senior Member, IEEE}, Carmine Clemente, \emph{Senior Member, IEEE}, \\ Silvia Liberata Ullo, \emph{Senior Member, IEEE}, and Danilo Orlando, \emph{Senior Member, IEEE},
\thanks{Filippo Biondi is with Italian Ministry of Defence. E-mail: biopippo@gmail.com.}
\thanks{Pia Addabbo is with the Universit\`a ``Giustino Fortunato'', 82100 Benevento, Italy. E-mail: p.addabbo@unifortunato.eu.}
\thanks{Carmine Clemente is with the University of Strathclyde, Department of Electronic and Electrical Engineering, 204 George Street, G1 1XW, Glasgow, Scotland.}
\thanks{Silvia Liberata Ullo is with the Universit\`a degli studi del Sannio, 82100 Benevento, Italy. E-mail: ullo@unisannio.it.}
\thanks{D. Orlando is with the Universit\`a degli Studi ``Niccol\`o Cusano'', 00166 Roma, Italy. E-mail: danilo.orlando@unicusano.it.}}

\markboth{submitted to IEEE Journal of Selected Topics on Geoscience and Remote Sensing}{}

\maketitle

\begin{abstract}

In this paper, authors propose a new procedure to provide a tool for monitoring critical infrastructures. Particularly, through the analysis of COSMO-SkyMed satellite data, a detailed 
and updated survey is provided, for monitoring the accelerating destabilization process of the Mosul dam, that represents the largest hydraulic facility of Iraq and is located on the Tigris river. 
The destructive potential of the wave that would be generated, in the event of the dam destruction, could have serious consequences.
If the concern for human lives comes first, the concern for cultural heritage protection is not negligible, since several archaeological sites are located around the Mosul dam.
The proposed procedure is an in-depth modal assessment based on the micro-motion estimation, through a Doppler sub-apertures tracking and a Multi-Chromatic Analysis (MCA).
The method is based initially on the Persistent Scatterers Interferometry (PSI) that is also discussed for completeness and validation.
The modal analysis has detected the presence of several areas of resonance that could mean the presence of cracks, and the results have shown that the dam is still in a strong destabilization. 
Moreover, the dam appears to be divided into two parts: the northern part is accelerating
rapidly while the southern part is decelerating and a main crack in this north-south junction is found. The estimated velocities through  the PS-InSAR technique show a good agreement with the GNSS in-situ measurements, resulting in a very high correlation coefficient and showing how the proposed procedure works efficiently.

\end{abstract}


\section{Introduction}
\IEEEPARstart{T}HE Mosul dam is located on the Tigris river and represents the largest hydraulic infrastructure in Iraq. The construction was started 1981 and completed five years later. Immediately and after impounding in 1986, seepage locations were recognized mainly due to the dissolution of gypsum \cite{adamo,5milillo}. 
With the rise of the self proclaimed Islamic State (IS), the dam has been under the control of IS in August 2014, for a short period of time. Maintenance and cement grouting operations have ceased since then and the spillways remained blocked, raising concerns about a possible dam failure, that could cause a catastrophic flooding and compromise the safety of more than 1.5 million people, living near the Tigris river.

\begin{figure}[htb!] 
	\centering
	\includegraphics[scale=1]{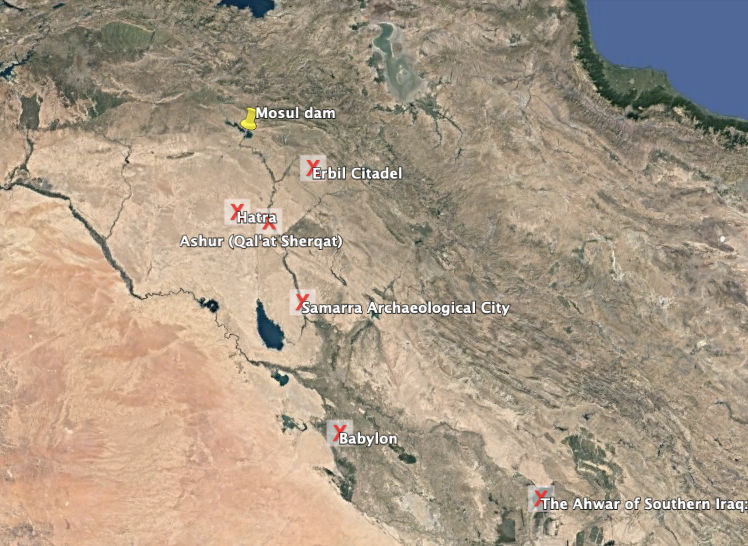}
	\caption{Unesco Iraq's Cultural Heritage sites: yellow pin indicates the Mosul dam location whereas red cross are the Unesco sites.}
	\label{unesco}
\end{figure}

If the concern for human lives comes first, the concern for cultural heritage protection is not negligible. The monitoring of the Mosul dam is, and, therefore its monitoring becomes of primary importance also for protecting the archaeological and historical heritage of Iraq, because its cultural heritage has suffered for decades from a range of destructive impacts. On the other hand, the relevance of the current state of Iraq within the worldwide cultural heritage is demonstrated by the inclusion of six sites on the UNESCO World Heritage List (Ashur, Babylon, Erbil Citadel, Hatra, Samarra, as shown in Figure \ref{unesco}), and of other eleven sites on the UNESCO World Heritage Tentative List. Another important initiative to be mentioned is the UNESCO-led project "Revive the Spirit of Mosul", underpinned by the February 2018 International Conference on the Reconstruction of Iraq, held in Kuwait City \cite{IraqCultHer}. 
All this highlights the international attention on the several archaeological sites located in Iraq, and focusing on the Mosul dam, its monitoring becomes of primary importance for the prevention of the many cultural heritage sites surrounding it.

Based upon the Persistent Scatterers Interferometry (PSI) \cite{PSI,Brcic2009DeltakWS,2006JB004763,2008GL034654}, the first satellite multi-sensor approach, to detect long time deformation displacement trends of the Mosul dam, is presented in \cite{Milillo_1,8128442}. 
In this work, the authors found rapid deformations during 2004-2010, which slowed in 2012-2014 and again accelerated since August 2014, when grouting operations stopped. The importance in using multiple sensors and different satellite systems was also remarked in \cite{Milillo_2}. With a focus on single systems, multiple satellite measurements can provide greater spatial coverage and temporal sampling, thereby enabling a better control on the interferometric decorrelation and a lower latency data access. These improvements lead to a more effective near real-time disaster monitoring, assessment, and response, as well as a greater ability to analyse dynamically changing physical processes. In \cite{Othman_1}, the Synthetic Aperture Radar (SAR) Persistent Scatterers Interferometry (PS-InSAR) is used to process Sentinel1-A SAR data for deformation monitoring of the Mosul dam, estimating a maximum deformation velocity of about 7.4 mm per year, at a longitudinal subsidence area extending over a length of 222 m along the dam axis. The mean subsidence velocity in this area is about 6.27 mm per year in the center of the dam. Subsidence rate shows also an inverse relationship with the reservoir water level.

An additional application of critical infrastructure monitoring using SAR is provided by \cite{1,ulloigarss}, where a methodology for the assessment of possible pre-failure bridge static deformations, based on DInSAR and PS-InSAR coherent processing, respectively,  is devised. A 15-year survey of the Morandi bridge is also given in \cite{1}, in the form of relative displacements across the structure, prior to its collapse on August 14th 2018. Static displacement maps are generated observing COSMO-SkyMed (CSK), Sentinel-1A/B and Envisat SAR data. Results reveal that the bridge was undergoing an increased amount of deformations over time, prior to its collapse. 

It is important to observe that, in order to carry out a detailed analysis of the infrastructure conditions, it is necessary to dynamically study them because of the variability over the time of its oscillations and structural variations. In fact, the infrastructure's movements are characterized by different values of displacement, velocity, and acceleration, distributed on scales that can also differ in orders of magnitude and frequencies. The technique developed in \cite{PSI} analyzes the movements that are of the order of millimeters per year and it is shown that, in order to detect them, it is necessary to process a long interferometric series of SAR data, that appears to be a critical point.

Another critical  factor of the SAR techniques, based on the differential interferometry, is the coherence between the analysed SAR images, which is of fundamental importance to obtain a reliable interferogram. In fact, this statistical parameter is very important because when it assumes low values, SAR interferometric fringes are chaotic and contain very little useful information. On the other hand, when the interferometric coherence magnitude is high (above 0.8), the interferometric fringes are very clear because they contain a significant amount of information. Unfortunately, SAR interferometric measurements are affected by several sources of decorrelation which limits the number of image pairs suitable for the displacement analysis, as also shown in a recent study on a DInSAR-based dam monitoring \cite{Pia_1}. 

One of the most important decorrelation sources is the Atmospheric Phase Screen (APS), but, this effect can be mitigated considering the APS estimation as shown in \cite{3}. 
Furthermore, interferogram formation requires images to be co-registered with an accuracy of better than a few tenths of a resolution cell to avoid significant loss of phase coherence. As for InSAR co-registration, a 2-D polynomial of low degree is usually chosen as warp function and the polynomial coefficients are estimated through least squares fit from the shifts measured on the image windows \cite{13,Pallotta}.

Beyond the InSAR and DInSAR techniques, the need for effectively identifying damages in complex structures has motivated the development of Structural Health Monitoring (SHM) paradigms. A comparatively recent development in SHM methods is the vibration-based damage identification. The basic premise of common vibration-based damage identification methods is that damage in a structure will alter the stiffness, mass or energy dissipation properties of the structure, which in turn will alter its measured dynamic response \cite{chen2018structural}. The work in \cite{doebling1996damage} investigated modal identification of large civil structures, such as bridges, under the ambient vibrational conditions. The results from the benchmark study show that the robustness of identified modes are judged by using their modal contributions to the measured vibration data. 

It is also worth to mention that some tracking techniques have been applied to monitor glacier movements, volcanic activities, and co-seismic tears in the solid earth \cite{7225114}, resulting from severe earthquakes, to address the technical defects and limitations of conventional Differential InSAR (DInSAR) techniques, namely their sparse coverage and the impact of dense vegetative cover \cite{14}. In the past, studies on offset tracking techniques, to measure slope movements, were dominated by optically sensed imagery from spaceborne or airborne platforms. Offset tracking techniques have been also more recently used for measuring very large Earth deformations  with low resolution SAR sensors \cite{15}.

This work proposes an in-depth modal assessment based on the micro-motion (m-m) estimation, through pixel tracking of
Doppler sub-apertures and Multi-Chromatic Analysis (MCA) \cite{7}. The main contributions are summarized as:
\begin{itemize}
\item The proposed method jointly exploits the modal analysis which is suitable to estimate the position of possible cracks and micro-motion estimation based upon MCA. A dynamic analysis is performed to measure the micro-motion movements that occur during a single SAR image acquisition time, usually consisting in a few seconds. By exploiting this technique, it is possible to estimate the vibration energy of the elastic bodies and therefore it is possible to
retrieve the exact position, both in space and in frequency, of
the vibrational anomalies, most likely associated to cracks in
the infrastructures \cite{pekau,OWOLABI}. 
\item The most important advantage
of the proposed method is that only a single image can be
used to perform the analysis, whereas existing methods need to
use several images.  In fact, also to enrich the experimental assessment, the displacement analysis, carried out by a PS-InSAR, is also here considered. It is worth to point out that the collection of a long temporal series of interferometric SAR observations is needed, in this case, to achieve reliable results. This kind of acquisition usually requires several months in order to obtain the minimum number of SAR images necessary to reliably separate, in the phase term, the atmospheric electromagnetic delay contribution from that due to displacement.
\item The proposed method could potentially either replace or work together with the in-situ sensors network, providing additive/complementary infrastructure health data. 
\end{itemize}
On the other hand, the main limitation of the proposed technique is related to the amount of the vibration energy detectable with respect to the noise. Fortunately, in the case discussed in this paper, we consider a very large infrastructure, which has very extended surfaces and, as a consequence, the Doppler perturbations of the received electromagnetic bursts can be easily detected.
Other limitations can also be related to sensor spatial resolution, the quality of the coregistrator, and, above all, the fact that measured targets must be stressed by external agents consisting of variable forces-fields over time.

Another limitation of the proposed approach could be represented by a particular polarization direction where vibrations are found to be unaffected. This happens when the polarization vector is parallel to the cross-slant-range direction. However, this can be a quite rare case because it is very difficult to observe vibrations perfectly linearly polarized with polarization vector parallel to the cross-slant-range.
On the contrary, purely linear polarized vibrations, parallel to the slant-range direction (i.e., compression waves), are the most effective to the SAR image, because moving targets more heavily perturbs Doppler history and are more easily detected as a time-domain displacement in the azimuth direction. 

However, if the radar does not have an efficient geometric view of the vibrations, there will be the possibility to scan in different geometries, for example, changing the observation direction from ''left/right-ascending'' to ''left/right-descending'' or changing the incidence angle.

The main difference between the present work and \cite{7} is that the vibrational model of the infrastructures has been linked to that of the radar micro-motion, and hence, that the application at hand represents a more difficult task with respect to estimating the ships' vibrations. For this reason, in order to make the approach effective, the sensitivity of the coregistrator has been enhanced.


The paper is organized according to the following scheme: the details of the signal processing technique are described in the methodology Section \ref{Methodology}. Section \ref{Experimental_Results} deals with the experimental results. The discussion of results is drawn in Section \ref{Discussion} and, finally Section \ref{Conclusions} concludes the paper.

\begin{figure}[htb!] 
	\centering
	\includegraphics[scale=0.33]{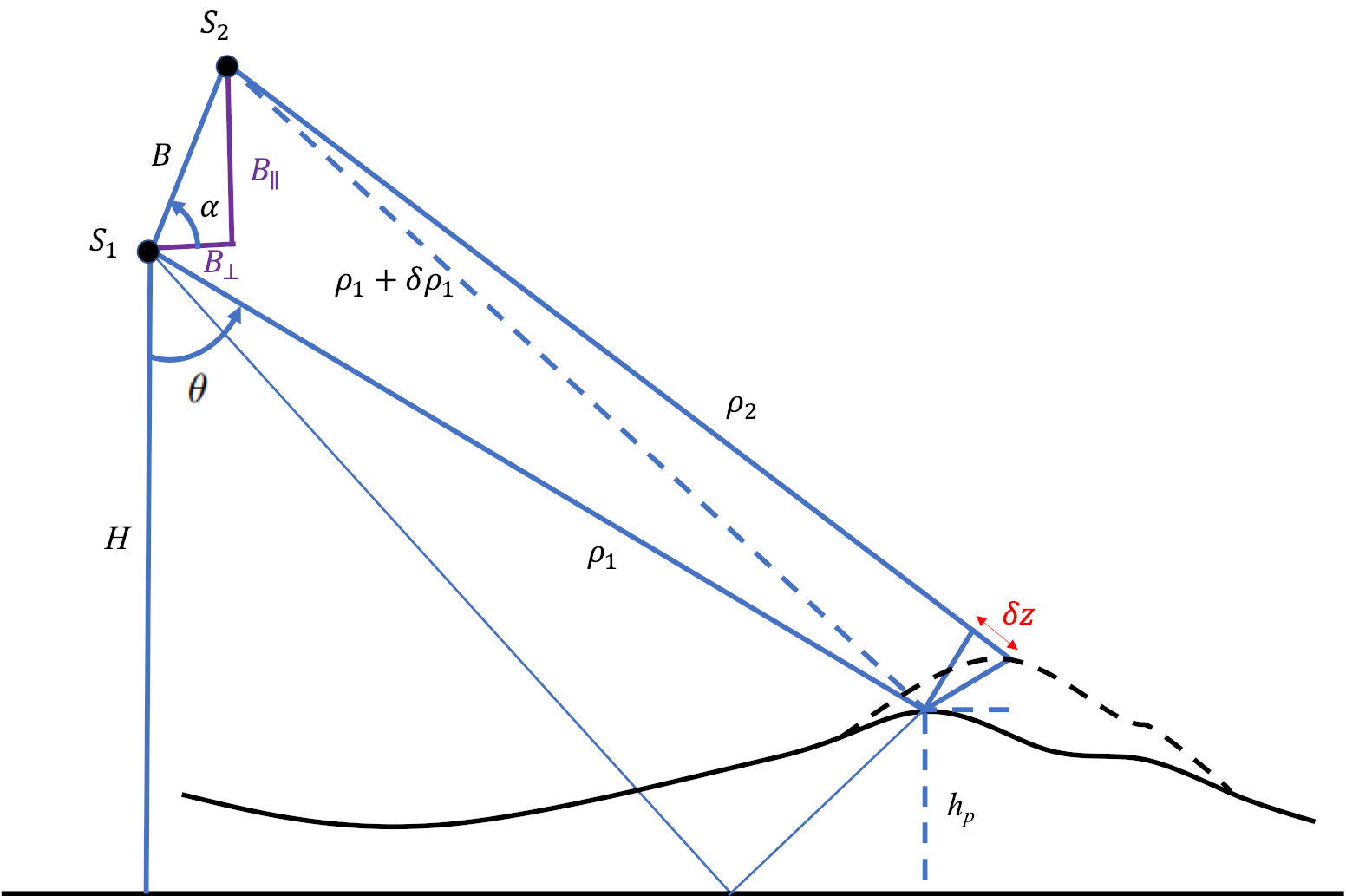}
	\caption{DInSAR acquisition geometry.}
	\label{Geometry_1}
\end{figure}

\section*{Notation}
In the sequel, vectors and matrices are denoted by boldface lower-case and upper-case letters, respectively. Let $u(t)$ be a function of $t$, then $\dot u(t)$ and $\ddot u(t)$ represent the first and the second derivative of $u$ with respect to the time $t$, respectively.
$f_X(x; p_1,...,p_n)$ denotes the Probability Density Function (pdf) of the continuous random variable $X$ with $n$ parameters $p_1, ... , p_n$.

\section{Methodology}\label{Methodology}

In this section, the proposed methodology is described in detail. The techniques used for the analysis are first introduced. Particularly, the method based on the PSI is discussed in Subsection \ref{PSI} whereas the proposed Modal Analysis is reported in Subsection \ref{M-M}.

\subsection{Persistent Scatterers Interferometry (PSI)}\label{PSI}

The observation geometry is depicted in Figure \ref{Geometry_1}: the SAR sensor is supposed to occupy two consecutive positions, namely $S_1$ and $S_2$, at the acquisition times $T_1$ and $T_2$, respectively. In the following, we suppose that data measured from $S_1$ represent the master image, whereas, data from $S_2$ belong to the slave one.
It is worth to notice that the selection of the master image is based on some hints, i.e., in order to maximize the expected interferometric coherence, based on the perpendicular and temporal baselines and the mean Doppler centroid frequency difference \cite{PSI}. 
A displacement event, represented by a dotted curve in Figure \ref{Geometry_1}, occurs among $T_1$ and $T_2$; $B$ is the baseline, $H$ is the height of the sensor in the position $S_1$ with respect to a reference surface and $h_p$ is the height of the terrain at time $T_1$; $\theta$ and $\alpha$ are the incidence angle and the master-slave sensor inclination angle respectively. The terms $\rho_1$ and $\rho_2$ represent the line of sight (LOS) target-sensor distances from $S_1$ and $S_2$, where $S_2$ is supposed to be at a distance $\rho_1+\delta \rho_1$ from the target at time instant $T_1$. Finally, $\delta_z$ is the LOS displacement, among $T_1$ and $T_2$, which has to be estimated.

The PSI is based on the processing of $N$ co-registrated SAR images generating $N-1$ differential interferograms with respect to a master image. 
The interferometric phase difference between each differential interferogram can be expressed in the form \cite{PSI}:
\begin {equation}
\Delta{\phi}=\Delta{\phi_{flat}}+\Delta{\phi_{elev}}+\Delta{\phi_{atmo}}+\Delta{\phi_{n}} + \Delta{\phi_{disp}},
\label{Global_Phase_1}
\end {equation}
with
\begin{itemize}
\item $\Delta{\phi_{flat}}$, the flat-Earth phase component due to the Earth curvature;
\item $\Delta{\phi_{elev}}$, the topographic contribution;
\item $\Delta{\phi_{atmo}}$, the atmospheric contribution due to the atmospheric humidity, temperature and pressure change between the two acquisitions;
\item $\Delta{\phi_{n}}$, the phase noise introduced by temporal change of the scatterers, different
look angle, and volume scattering;
\item $\Delta{\phi_{disp}}$,  the phase displacement.
\end{itemize}

The aim of the InSAR technique is to isolate the phase displacement contribution. 
The flat Earth and topographic phase components can be easily removed based on geometric considerations, whereas the atmospheric contribution removal does not represent an easy task.

In order to separate the phase displacement $\Delta{\phi_{disp}}$ from the parameter $\Delta{\phi_{atmo}}$, the approach presented in \cite{PSI} requests a long temporal series of interferograms in order to separate the displacement phase component from the APS. 
As an alternative solution, we propose to estimate the parameter $\Delta\phi_{atmo}$ by considering the algorithm proposed in \cite{3}.

We select the candidate PSs analyzing the time-series amplitude values of each pixel in the area of interest and looking for stable scatterers. 
 According to \cite{PSI}, the amplitude $A$ of a SAR image is supposed to follow the Rice distribution as	
\begin {eqnarray}
f_A(a;\nu,\sigma)=\frac{a}{\sigma^2}\exp \left(\frac{-(a^2+\nu^2)}{2\sigma^2}\right)I_0 \left(\frac{a\nu}{\sigma^2}\right),
\label{Rice_1}
\end {eqnarray}
where the parameters $\nu$ and $\sigma$ are the observed energy and the pixel distribution variance of each single-look-complex (SLC) image components, whereas, $I_0$ is the first kind modified Bessel function with shape parameter $\frac{v^2}{2\sigma^2}$. 
The amplitude dispersion index, $D_A$, is defined as \cite{PSI}
\begin{equation}
D_A= \frac{\sigma}{\nu},
\label{DA}
\end{equation}
which represents a measure of the amplitude stability. A threshold is experimentally found to select a subset of pixels within the SAR image that are candidate PSs.

\subsection{Single-Image Modal Analysis using SAR Micro-Motion}\label{M-M}

The aim of this subsection is twofold. First, we introduce the model to describe the vibrations generated by a distributed body. Then, we incorporate the latter into a procedure to estimate the vibrations of a distributed structure by means of radar data.

\paragraph{Vibrational Model of Infrastructures} \label{Oscillations_1}
The model of motion for the linear damped forced vibration of a structural dynamic system with a total number of $M$ degrees of freedom can be expressed as \cite{chen2018structural}
\begin {eqnarray}\label{eq_1}
\mathbf{M}\ddot{\mathbf{u}}(t) + \mathbf{C} \dot{\mathbf{u}}(t) + \mathbf{K} \mathbf{u}(t) = \mathbf{f}(t).
\end {eqnarray}
where:
\begin{itemize}
\item $\mathbf{f}(t)$ is the external force vector applied to the system over the time $t$;
\item $\mathbf{u}(t)$, $\dot{\mathbf{u}}(t)$ and $\ddot{\mathbf{u}}(t)$ are the nodal displacement, the velocity, and the acceleration vectors, respectively;
\item $\mathbf{M}$, $\mathbf{C}$ and $\mathbf{K}$ are the $M \times M$ global mass, damping and stiffness matrices of the dynamic system, respectively.
\end{itemize}
In forced harmonic vibration, if $\mathbf{f}(t)=\mathbf{f}e^{j\omega t}$ with a driving frequency $\omega$, Equation \eqref{eq_1} becomes
\begin {eqnarray}\label{eq_2}
\left(-\omega^2 \mathbf{M} + j\omega \mathbf{C} +\mathbf{K}\right) \mathbf{y} =\mathbf{f},
\end {eqnarray}
where $\mathbf{y}$ is the harmonic displacement vector
\begin {eqnarray}\label{eq_3}
\mathbf{y}=\mathbf{H}(\omega)\mathbf{f} \end {eqnarray}
with $\mathbf{H}(\omega)=\left(-\omega^2 \mathbf{M} + j\omega \mathbf{C} + \mathbf{K}\right)^{-1}$, the frequency response that represents the dynamic flexibility of the structural system.
The dynamic stiffness $\mathbf{Z}(\omega)$ can be defined as the inverse of the frequency response  \cite{chen2018structural}
\begin {eqnarray}\label{eq_4}
\mathbf{Z}(\omega)=\mathbf{H}(\omega)^{-1} = \left(-\omega^2 \mathbf{M}+j\omega\mathbf{C}+\mathbf{K}\right).
\end {eqnarray}

In a structural dynamic testing, the frequency response function is obtained by measuring the system responses at different locations, and, the displacement vector $\mathbf{y}$ in \eqref{eq_2} is given by  an appropriate measurement tool. In this paper, we will use the radar sensor to estimate $\mathbf{y}$ and the details of the proposed estimation procedure are given in the next subsection.

\paragraph{Radar Micro-Motion Model}\label{Oscillations_2}
The dynamic displacement estimation problem of a target performed by radar is here discussed. Standard SAR-processing methods are based upon the assumption of a static scene. If targets are moving, their positions in the SAR-image are shifted in azimuth and range-azimuth and a defocusing may occur. According to \cite{7}, the target defocusing, $\delta R(t)$ can be expressed as
\begin {eqnarray}\label{eq_new20}
\delta R(t) = \frac{2 L v_x(t)}{v_p \sin \gamma}.
\end {eqnarray}
where $\gamma$ is the the angle between the velocity vector and the line of sight,  $v_x(t)$ is the azimuth component target velocity during the time of measurement $\frac{L}{v_p}$, with $L$ the synthetic aperture and $v_p$ the platform velocity. 
Moreover, the motion leads to smearing effects on the focused signals in both range and cross-range directions. Moving targets produce significant position errors in azimuth due to the 
target range velocity component. A pure displacement along the azimuth direction returns a pure defocusing along the same direction. All these effects are detected by two-dimensional normalized cross-correlation and tracked during the SAR acquisition orbital time by the sub-pixel coregistration process. In the following, the process is explained in details.

\begin{figure}
	\centering
	\begin{subfigure}
		\centering
		\includegraphics[width=7.5cm,height=4.8cm]{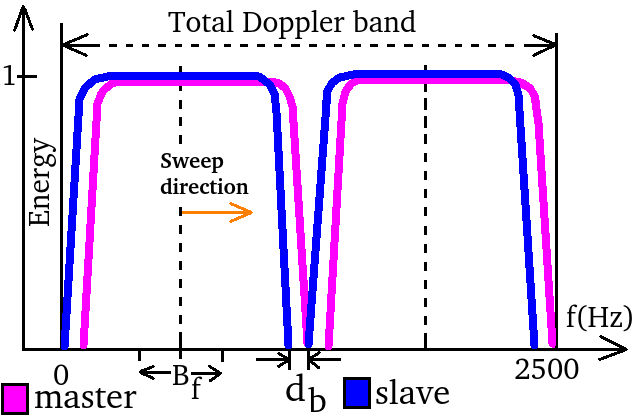}
	\end{subfigure}\quad
	\caption{Frequency allocation plane strategy.}
	\label{Frequency_Allocation_Plane}
\end{figure}

Sub-Pixel Offset Tracking (SPOT) is a relevant technique to measure large-scale ground displacements in both range and azimuth directions and it is complementary to differential interferometric SAR and persistent scatterers interferometry in the case of radar phase information instability \cite{7}. In this paper, we apply the above pixel tracking technique to the single stripmap image instead of using multi-temporal interferometric images. Specifically, the single observation is divided into several Doppler sub-apertures in order to investigate the fastest displacements due to moving targets (note that the acquisition duration is in the order of a few seconds). Thus, we focus on the vibrations of some specific points estimating the dynamic displacements of the scatterers during the Doppler sub-aperture history. In accordance with the frequency subdivision strategy of Figures \ref{Frequency_Allocation_Plane}, we observe the offset trend by computing the normalized cross correlation after partitioning the image into small patches. The estimation procedure consists in shifting the master for each Doppler event and calculating the correlation between adjacent Doppler sub-apertures, according to a small-frequency baseline strategy. 

The offset components of the sub-pixel normalized cross-correlation, according to \cite{7} and \cite{6} are described by the complex parameter ${{\mathbf{{D}}}^{i,j}_{\textrm{tot}_{({c,D})}}}$ referring to the total displacement, which is estimated by
\begin {eqnarray}\label{eq_s1}
\begin{aligned}
	{{\mathbf{{D}}}^{i,j}_{\textrm{tot}_{({c,D})}}}&=  {{\mathbf{{D}}}^{i,j}_{\textrm{displ}_{({c,D})}}}+{{\mathbf{{D}}}^{i,j}_{\textrm{topo}_{({c,D})}}}+{{\mathbf{{D}}}^{i,j}_{\textrm{orbit}_{({c,D})}}}&&\\ 
	&+{{\mathbf{{D}}}^{i,j}_{\textrm{control}_{({c,D})}}}+{{\mathbf{{D}}}^{i,j}_{\textrm{atmosphere}_{({c,D})}}}+{{\mathbf{{D}}}^{i,j}_{\textrm{noise}_{({c,D})}}},&&\\
	&i=1, \dots, G_D, j=1, \dots, G_t,&&
\end{aligned}
\end {eqnarray}
where:
\begin{itemize}
	\item ${{\mathbf{{D}}}^{i,j}_{\textrm{displ}_{({c,D})}}}$ is the offset component generated by the infrasctructure vibrational trend and detected as a sub-pixel misalignment existing between the first SAR image (master) and the $i-$th slave Doppler sub-aperture;
	\item $\mathbf{{D}}^{i,j}_{\textrm{topo}_{({c,D})}}$ is the offset component generated by the earth displacement when located on highly sloped terrain;
	\item ${{\mathbf{{D}}}^{i,j}_{\textrm{orbit}_{({c,D})}}}$ is the offset caused by residual errors of the satellite orbits;
	\item ${{\mathbf{{D}}}^{i,j}_{\textrm{control}_{({c,D})}}}$ is the offset component generated by general attitude and control errors of the flying satellite trajectory;
	\item ${{\mathbf{{D}}}^{i,j}_{\textrm{atmosphere}_{({c,D})}}}$ and ${{\mathbf{{D}}}^{i,j}_{\textrm{noise}_{({c,D})}}}$ are the contributions generated by the electromagnetic aberrations due to atmosphere parameters space and time variations and general disturbances due to thermal and quantization noise, respectively.
\end{itemize}

\begin{figure}[htb!]
	\centering
	\begin{subfigure}
		\centering
		\includegraphics[width=7cm,height=6.0cm]{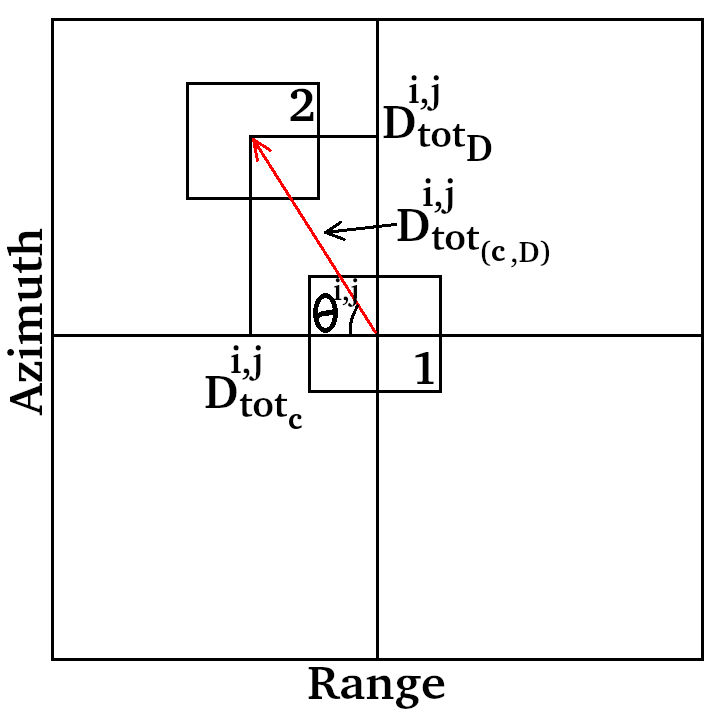}
	\end{subfigure}\quad
	\caption{Schematic representation of isolated pixels with a certain shift due to space displacement.}\label{Schema_Pixel_Tracking}
\end{figure}

This formula is a general case where displacement exists in both range-azimuth dimensions and evolving in time. It is also worth to notice that vibrations could be generated by various forces, both of natural and of anthropogenic nature, with the predominance of the one with respect to the other \cite{1143684}.
However, we assume that the main contribution of ${\mathbf{{D}}}^{i,j}_{\textrm{tot}_{({c,D})}}$ is due to the reprojection of these cumulative effects in the range/azimuth domain.


Figure \ref{Schema_Pixel_Tracking} is a schematic representation of the parameters estimated by the coregistration procedure. The square number one is a focused pixel of the master image and the square number two is the same pixel but located on the slave image. The parameters ${{\mathbf{{D}}}^{i,j}_{\textrm{tot}_{c}}}$ and $\theta^{i,j}$ are the distance between the master and slave pixel centers and the angle respect to the horizontal axis respectively. Distances are estimated by two-dimensional cross-correlation. In the present case, since the shift due to vibrations occurs on the range-azimuth plane, the parameter $\theta^{i,j}$ is an harmonic parameter.

As stated before, Figure \ref{Frequency_Allocation_Plane} represents the frequency allocation plane strategy used to estimate the m-m. In order to analyze the vibrational anomalies, detectable using \eqref{eq_new20}, we induce a frequency variation of the focusing band in azimuth.  
The estimated shifts ${\mathbf{{D}}}^{i,j}_{\textrm{tot}_{({c,D})}}$ are now available and used in conjuction with \eqref{eq_3} as follows
\begin {eqnarray}\label{eq_s2}
&\sum\limits_{i=1}^{G_D}\sum\limits_{j=1}^{G_t}\sum\limits_{l=1}^{N_c-1}\sum\limits_{m=1}^{N_D-1} {{\mathbf{{D}}}^{{i,j}^{\{l,m\}}}_{\textrm{tot}_{({c,D})}}}\\ \nonumber
&=\sum\limits_{j=1}^{G_t}\sum\limits_{l=1}^{N_c-1}\sum\limits_{m=1}^{N_D-1}\mathbf{H}^{\{l,m\}}_j(\omega)\mathbf{f}^{\{l,m\}}_j,
\end {eqnarray}
where $N_c$ is the number of image's pixels along the range, $N_D$ is the pixels' number along the range, and $G_D$ and $G_t$ are the lengths of the particular frequency and temporal series, respectively. 
 Equation \eqref{eq_s2} gives an operational solution for SAR to the vibrational model \eqref{eq_3}.
 
By summarizing, the whole processing chain is composed by 8 functional blocks as shown in Figure 5. Specifically, they perform the following operations: 
\begin{itemize}
\item The first step consists in the selection of a single SAR raw image observing the dam. 
\item The bi-dimensional spectrum of the selected image is calculated via a 2D-DFT in the second step. 
\item The third block consists in a band-pass filtering according to the small-frequency baseline strategy of Figure 3. The output of the third block consists in multiple images centered at the consecutive different Doppler frequencies.
\item The sub-aperture images, re-transformed into the raw data domain via an inverse DFT (functional block 4), represent the input for the successive SAR processing  blocks (5-8). 
\item The fifth block is devoted to two-dimensional focusing \cite{cumming}.
\item The coregistration process, which is performed by block 6, consists in aligning the pixels of any slave image with respect to the corresponding pixels of the master image. The alignment process is very precise and occurs at the sub-pixel level (the coregistration  parameters are given in Table I).
\item In the Pixel Tracking (functional block 7), the total displacement in (9) is computed.
\item In the last block, the Modal Analysis solves Equation (10).
\end{itemize}


\begin{figure*}[htb!] 
	\centering
	\includegraphics[scale=0.45]{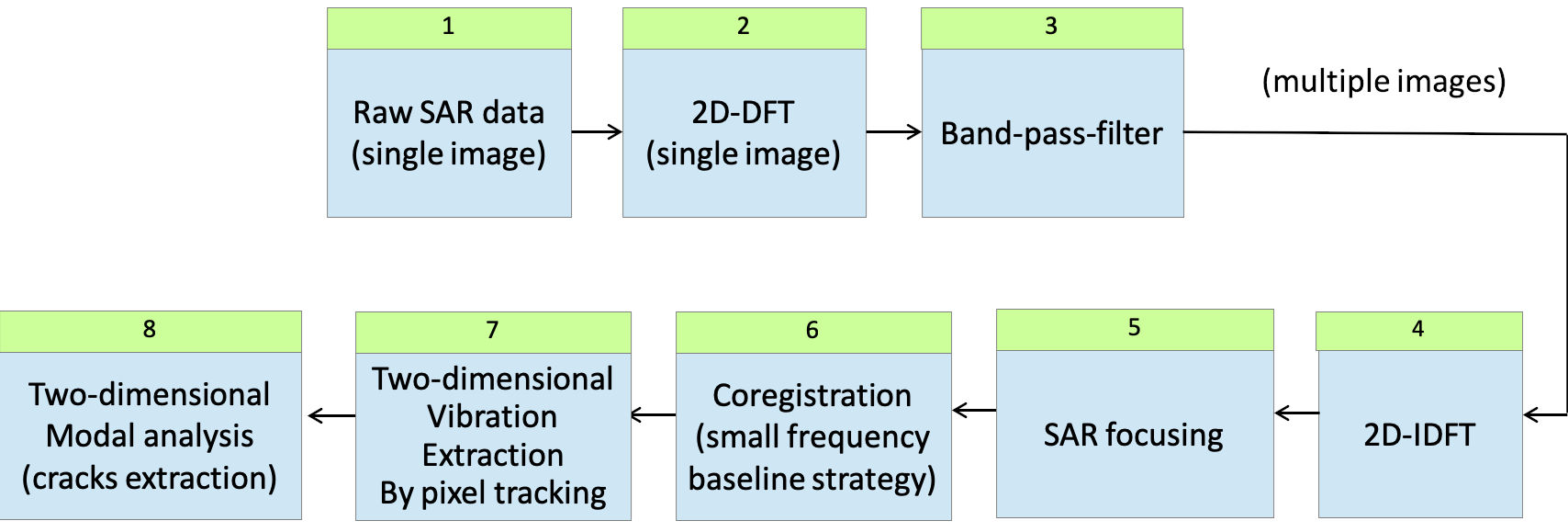}
	\caption{Functional Block scheme of the processing chain used for the Modal Analysis.}
	\label{scheme}
\end{figure*}

It is also worth to conclude this section with the two following important  practical hints:
\begin{itemize}
\item If the radar does not have an efficient geometric view of the vibrations, there will always be the possibility to scan in different geometries, for example, changing the observation direction from ''left/right-ascending'' to ''left/right-descending'' or changing the incidence angle.
\item The proposed technique may not work when the target is small, which means it occupies a few or even a resolution cell. Another challenging case is when the target is not very visible from the radar because the noise maintains a certain predominance over the signal. Problems may also occur when the target, although large and well visible, is lying well attached on very stable ground and when the target, although large and well visible, is not subjected to any force that destabilizes its staticity. It should be noted that when a target is subjected to forces, although large but sporadically distributed over time, could not be captured by a single SAR observation. In fact, let us recall that the radar generates SAR images that is an approximated representation of a situation frozen during the observation time interval that consists in a few seconds, for the stripmap mode, or a few tens of seconds, in the spotlight case.
\end{itemize}

\section{Experimental Results}\label{Experimental_Results}
\begin{figure}[htb!] 
	\centering
	\includegraphics[width=7.5cm,height=6.0cm]{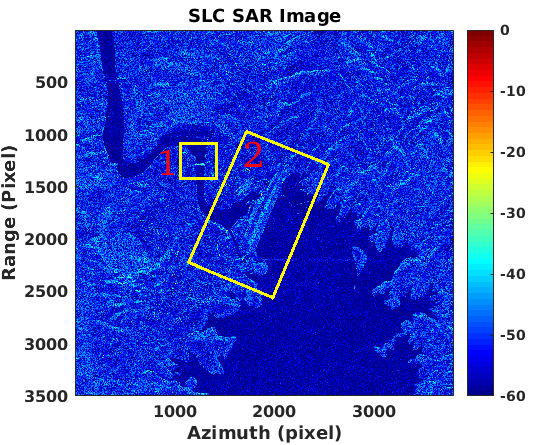}
	\caption{SLC SAR image case studies.}
	\label{SLC_all}
\end{figure}

The experimental analysis focuses on two case studies. The geographical location  of both case studies concerns the Mosul dam reservoir in Iraq. The first case study is the analysis of a bridge, located near the dam, that crosses the Tigris river. This infrastructure is visible inside the yellow labelled ``1'' of Figure \ref{SLC_all}.  The second case study is more exhaustive and deals with the dam as shown by the yellow box ``2'' in the figure.

For the selection of the candidate PSs, we have considered all the pixel with an inverse amplitude stability of 
$0.8$ and the coregistration parameters used in the analysis are reported in Table \ref{Tab_2}.
The other processing parameters are reported into Table \ref{Tab_1}. It is possible to notice that both case studies are analysed through Modal and PS-InSAR techniques. For the last method a number of 83 SAR images are needed with an observations' temporal extension of more than $6$ years.

Furthermore, from the PS-InSAR displacements' estimation, it is possible to retrieve both velocities and accelerations, evaluated as the first and the second order finite differences approximation respectively. 
Conversely, through the Modal analysis, the vibration energy is estimated.

\begin{table}[htb!]
\caption{Coregistration parameters}
\centering\label{Tab_2}
\begin{tabular}{ccc}
\toprule
\textbf{Parameter}	& \textbf{Value}\\
\midrule
Initial shifts & Coarse cross-correlation \\
Number of points   & 1450 \\
Correlation threshold & 0.8 \\
Oversampling factor  & 1200 \\
search pixel window  & $10 \times 10$ pixels \\
Points skimming & 30 \\
Use of DEM & No \\
Doppler Centr. Est. Strategy & Polynomials \\	
\bottomrule	
\end{tabular}
\end{table}

\begin{figure*}[htb!] 
	\centering
	\includegraphics[width=15.0cm,height=7cm]{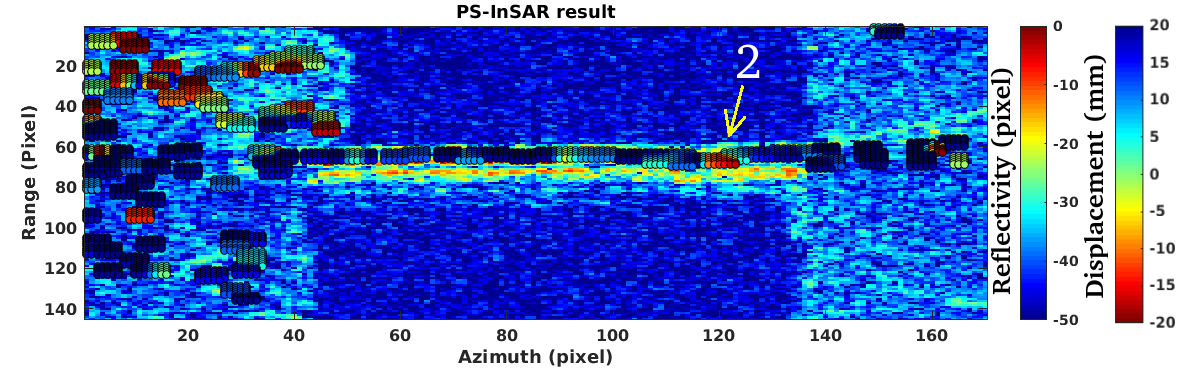}
	\caption{Displacement map of Case Study 1-a (PS-InSAR): red circles indicate displacement away the sensor and blue circles indicate the same effects but towards the sensor.}
	\label{Bridge_1_1}
\end{figure*}

\begin{figure*}[htb!] 
	\centering
	\includegraphics[width=15.0cm,height=7cm]{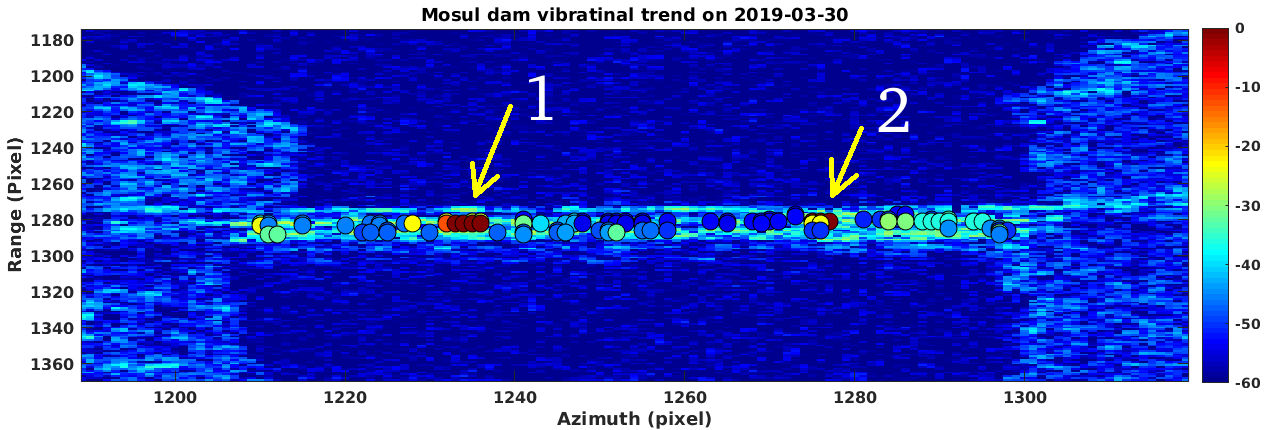}
	\caption{The vibrational map of Case Study 1-b (Modal Analysis): the circles indicate the vibrational values for some selected pixels with high energy reflectivity. Arrows "1" and "2" indicate two vibrational anomalies.}
	\label{Bridge_1}
\end{figure*}

\begin{figure*}[htb!] 
	\centering
	\includegraphics[width=14cm,height=6cm]{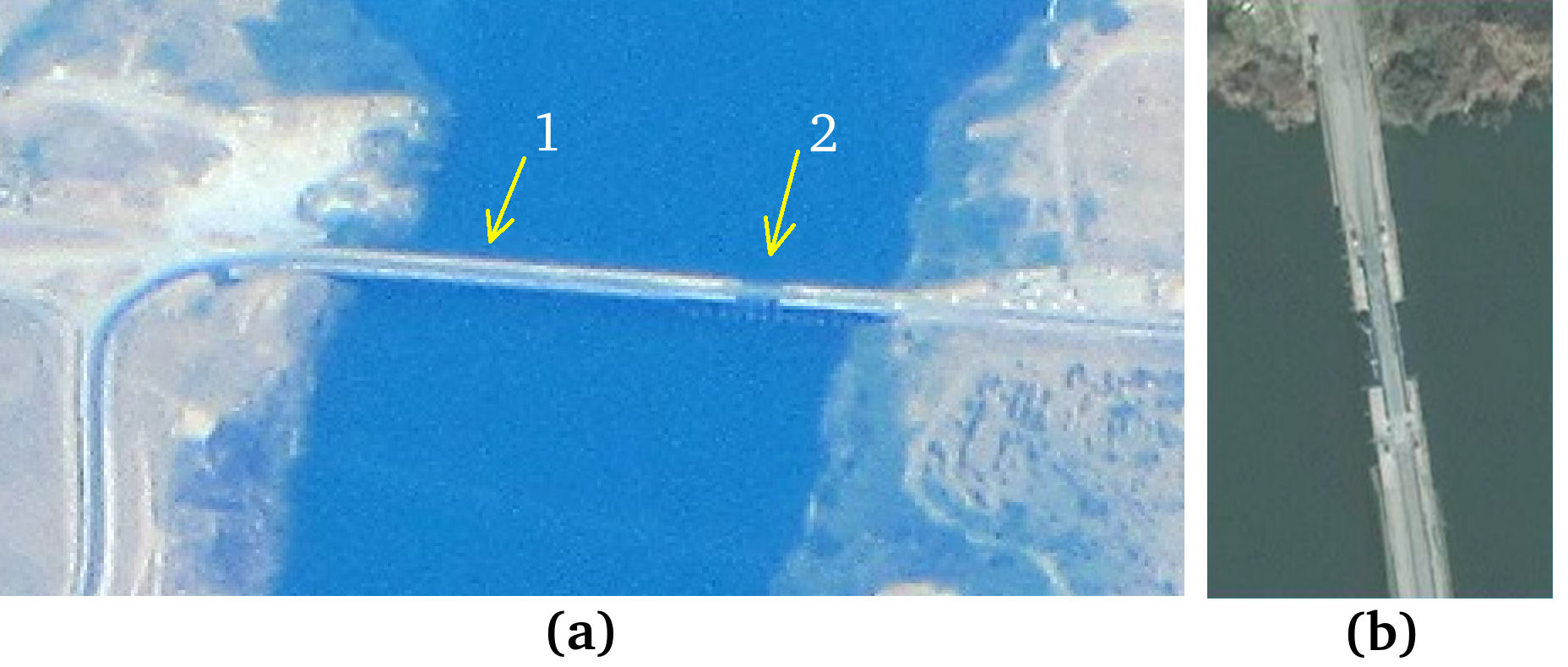}
	\caption{The optical images of Case Study 1, In Subfigure (a) Arrows "1" and "2" indicate the positions of the two vibrational anomalies. Subfigure (b) is a zoomed version around the arrow labelled "2".}
	\label{Bridge_2}
\end{figure*}

\begin{figure}[htb!] 
	\centering
	\includegraphics[width=7.0cm,height=5.5cm]{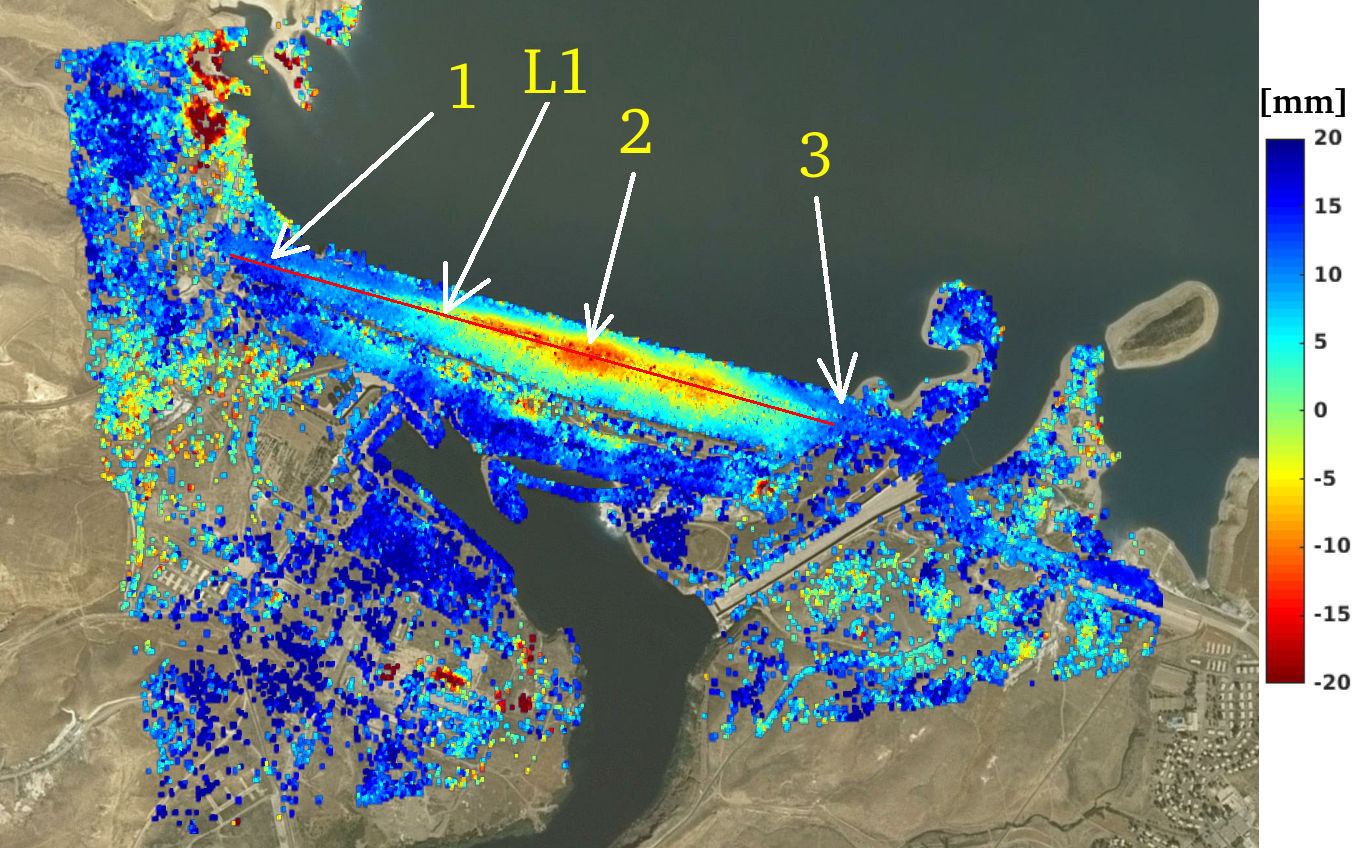}
	\caption{Displacement map (case study 2-a): numbers indicate the principal measurement poins where the displacement history over time has been considered. L1 represents a line where all displacements, velocities and accelerations have been studied.}
	\label{Cumulative_Displacement_all}
\end{figure}

\begin{figure}[htb!] 
	\centering
	\includegraphics[width=7.0cm,height=5.5cm]{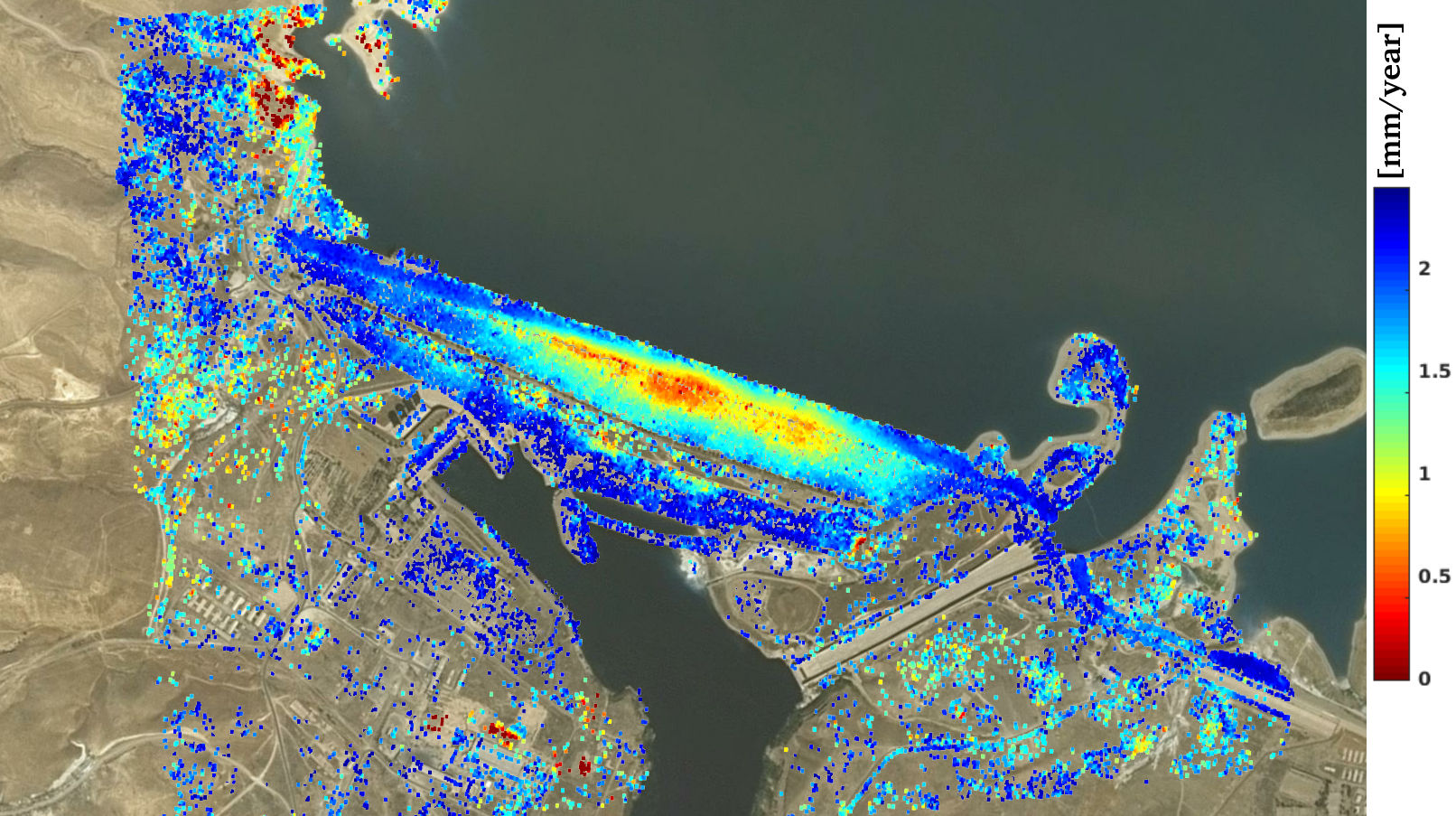}
	\caption{Velocity map (case study 2-a): first-order displacement derivative.}
	\label{Cumulative_Velocities}
\end{figure}

\begin{figure}[htb!] 
	\centering
	\includegraphics[width=7.0cm,height=5.5cm]{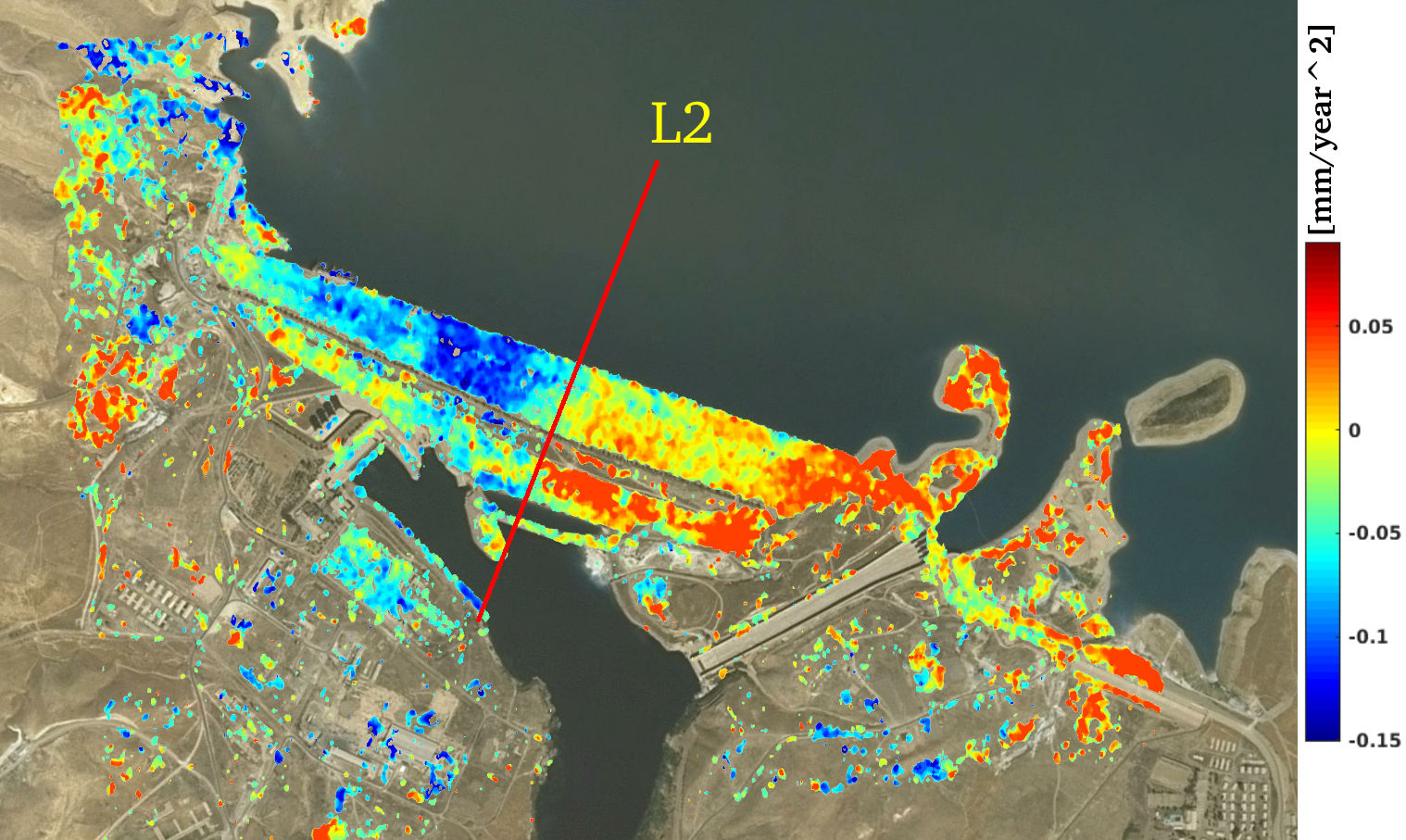}
	\caption{Acceleration map (case study 2-a): second-order displacement derivative. L2 represents a line of accelerations' inversion. }
	\label{Cumulative_Accellerations_All}
\end{figure}

\begin{figure}[htb!] 
	\centering
	\includegraphics[width=7.0cm,height=5.5cm]{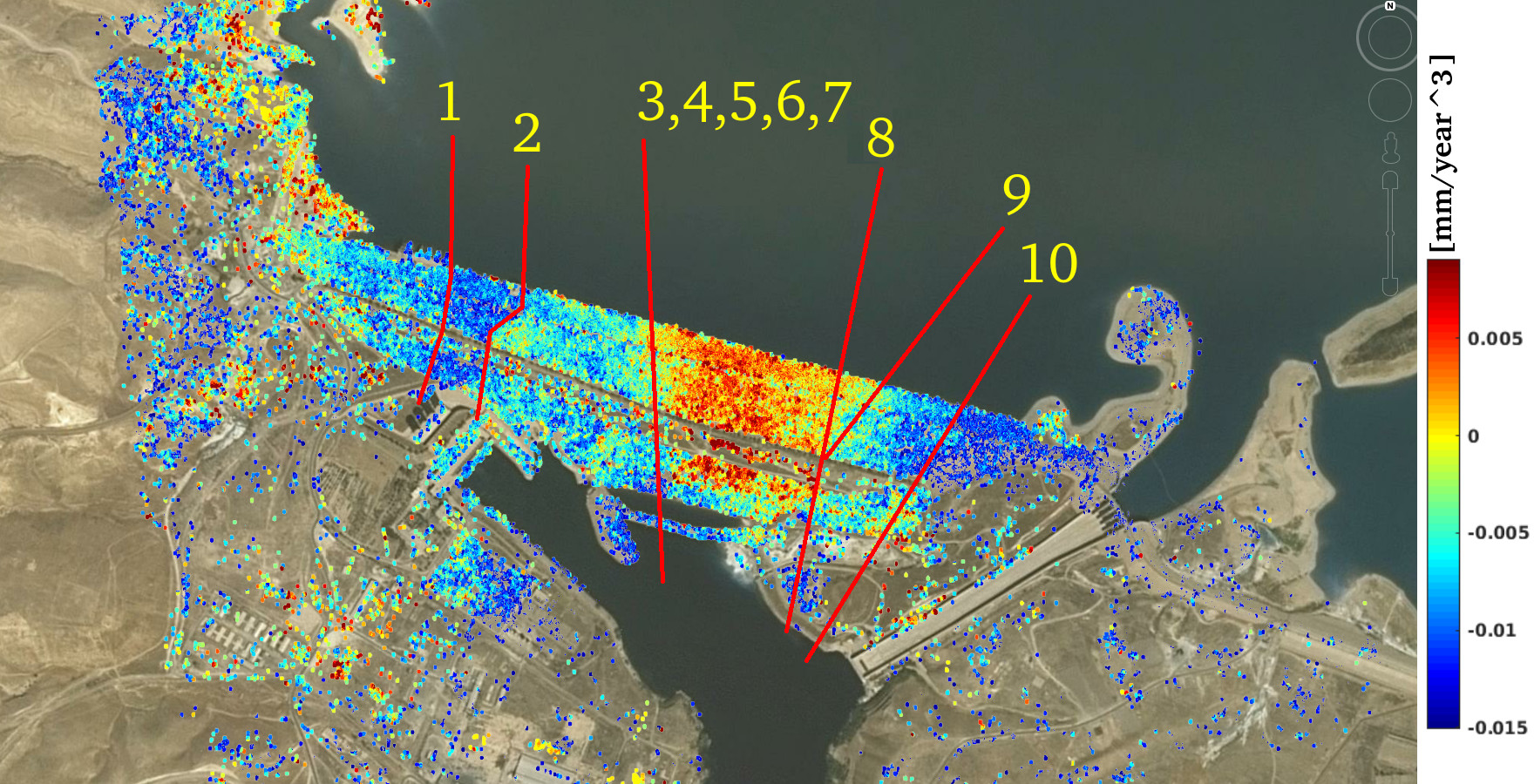}
	\caption{First order variations of acceleration (case study 2-a): third-order displacement derivative.}
	\label{Cumulative_Accellerations_Var_1}
\end{figure}

\begin{figure}[htb!] 
	\centering
	\includegraphics[width=7.0cm,height=5.5cm]{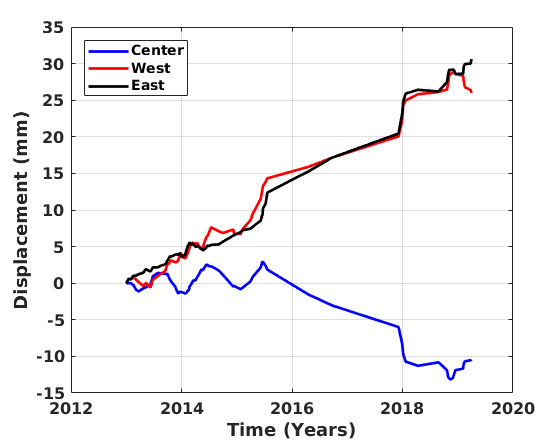}
	\caption{Displacement history versus time measured on point 1 (red line), point 2 (blue line) and point 3 (black line). All measurement points are the one represented in Figure \ref{Cumulative_Displacement_all}.}
	\label{Displacement_1}
\end{figure}

\begin{figure}[htb!] 
	\centering
	\includegraphics[width=8.0cm,height=6.5cm]{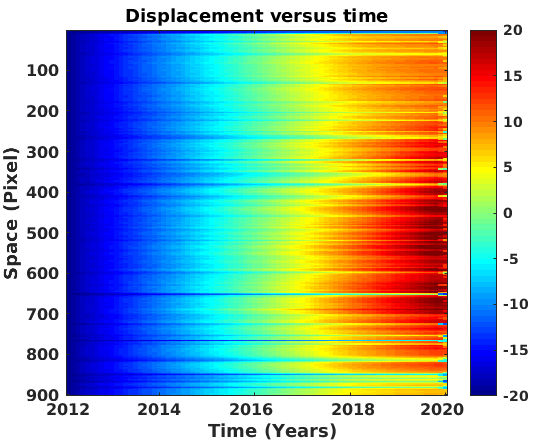}
	\caption{Displacement versus time (case study 2-a): third-order displacement derivative. }
	\label{Line_Displacement}
\end{figure}

\begin{figure}[htb!] 
	\centering
	\includegraphics[width=8.0cm,height=6.5cm]{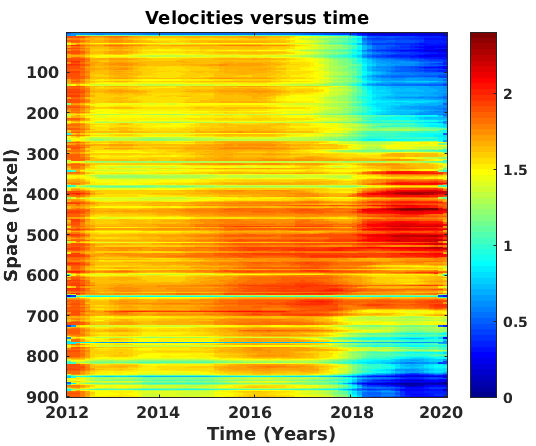}
	\caption{Velocities versus time (case study 2-a): first-order displacement derivative.}
	\label{Line_Velocity}
\end{figure}

\begin{figure}[htb!] 
	\centering
	\includegraphics[width=8.0cm,height=6.5cm]{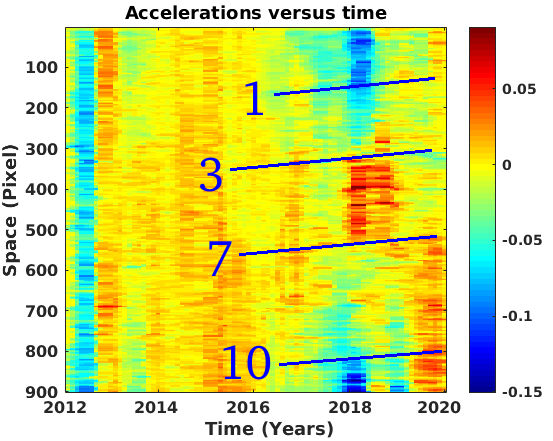}
	\caption{Accelerations versus time (case study 2-a): second-order displacement derivative. Numbers indicate the retrieved cracks.}
	\label{Line_Acceleration}
\end{figure}

\begin{figure}[htb!]
	\centering
	\includegraphics[width=8.5cm,height=6.5cm]{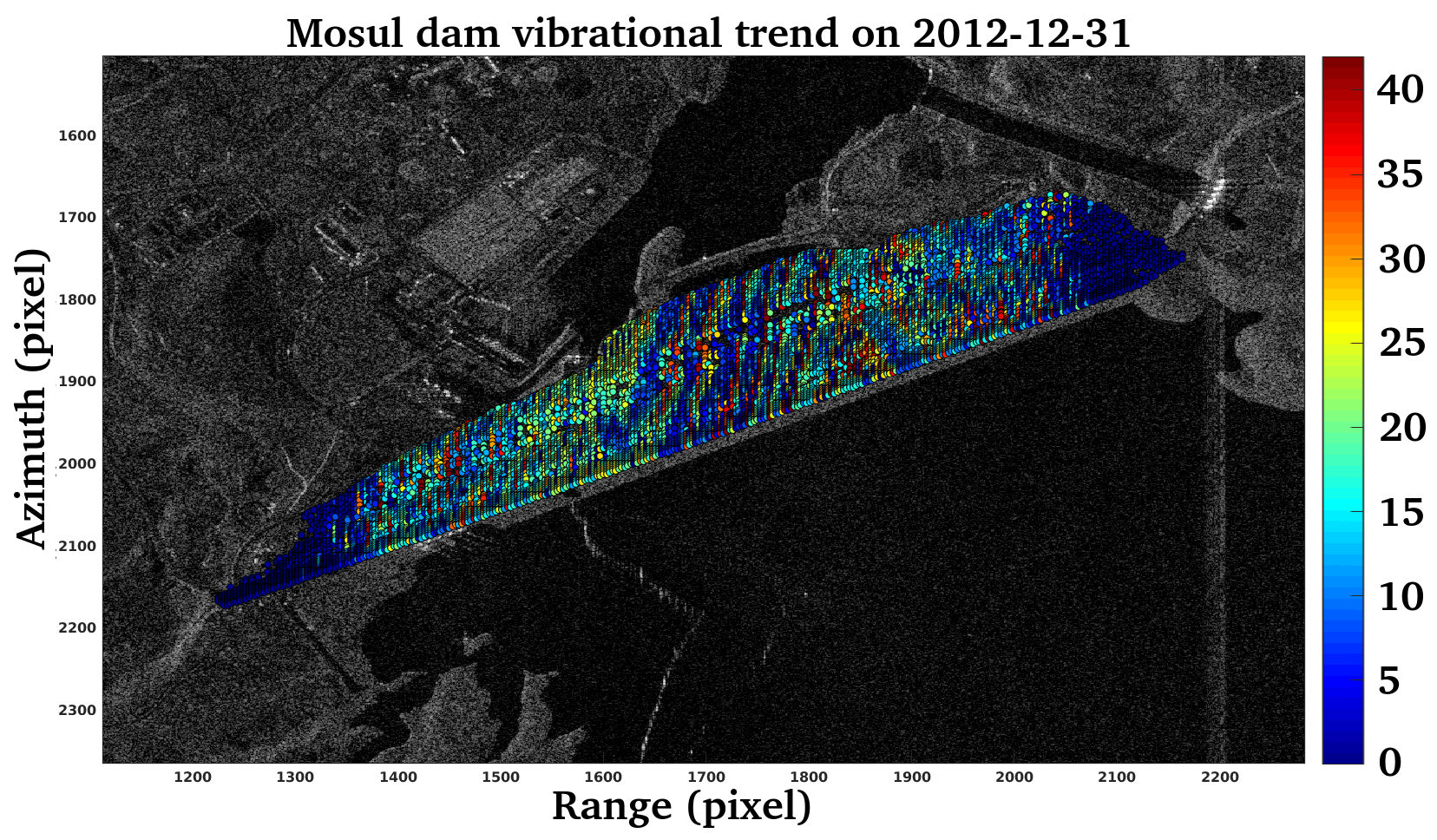}
	\caption{Vibrations estimated on 31 December 2012 (case study 2-b).}
	\label{Vibrazioni_1}
\end{figure}

\begin{figure}[htb!]
	\centering
	\includegraphics[width=8.5cm,height=6.5cm]{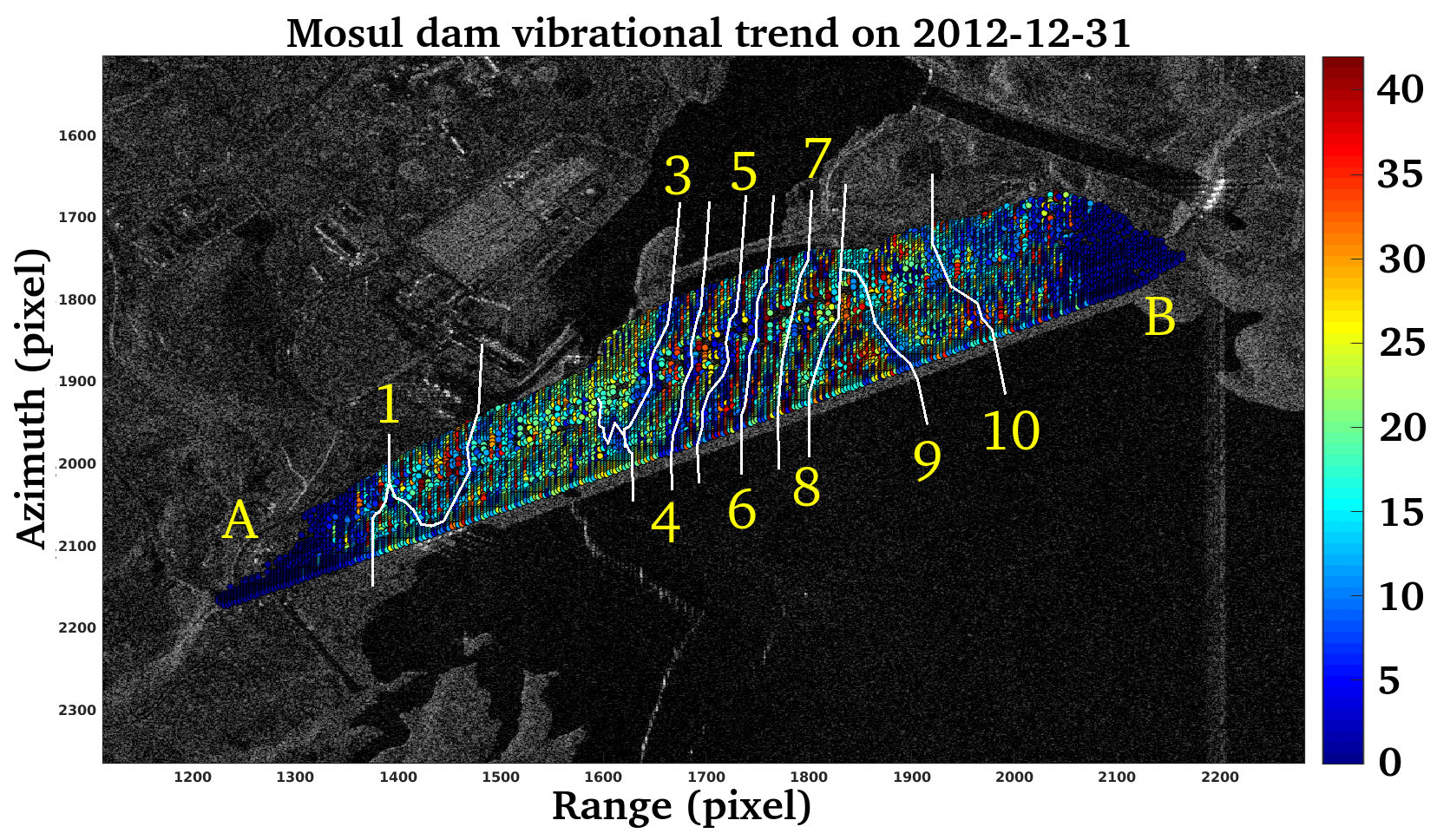}
	\caption{Vibrations estimated on the 31 December 2012 (case study 2-b). Numbers indicate the retrieved cracks. Letters represent the dam's edges. }
	\label{Vibrazioni_2}
\end{figure}

\begin{figure}[htb!]
	\centering
	\includegraphics[width=8.5cm,height=6.5cm]{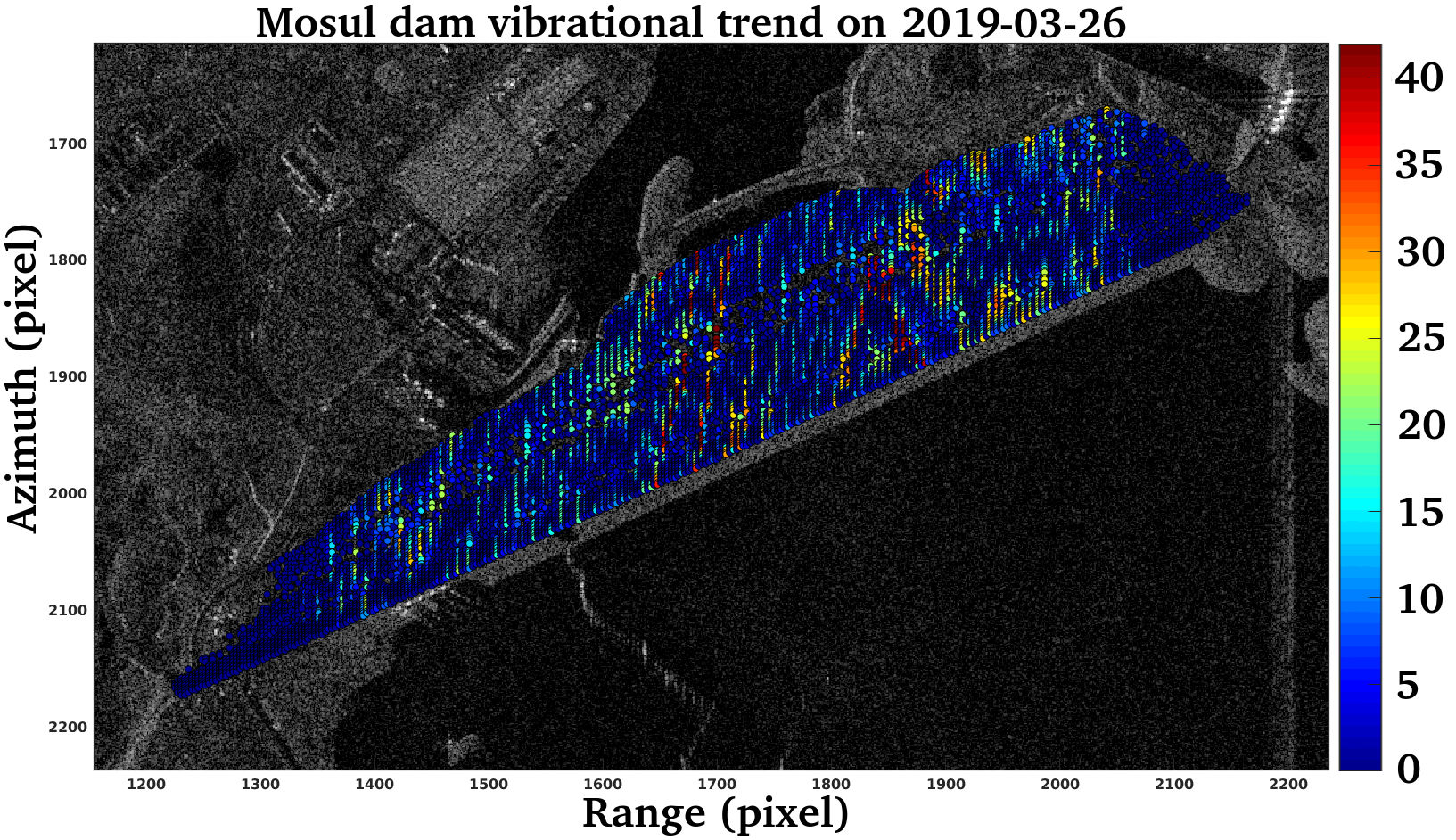}
	\caption{Vibrations estimated on the 26 March 2019 (case study 2-c).}
	\label{Vibrazioni_3}
\end{figure}

\begin{figure}[htb!]
	\centering
	\includegraphics[width=8.5cm,height=6.5cm]{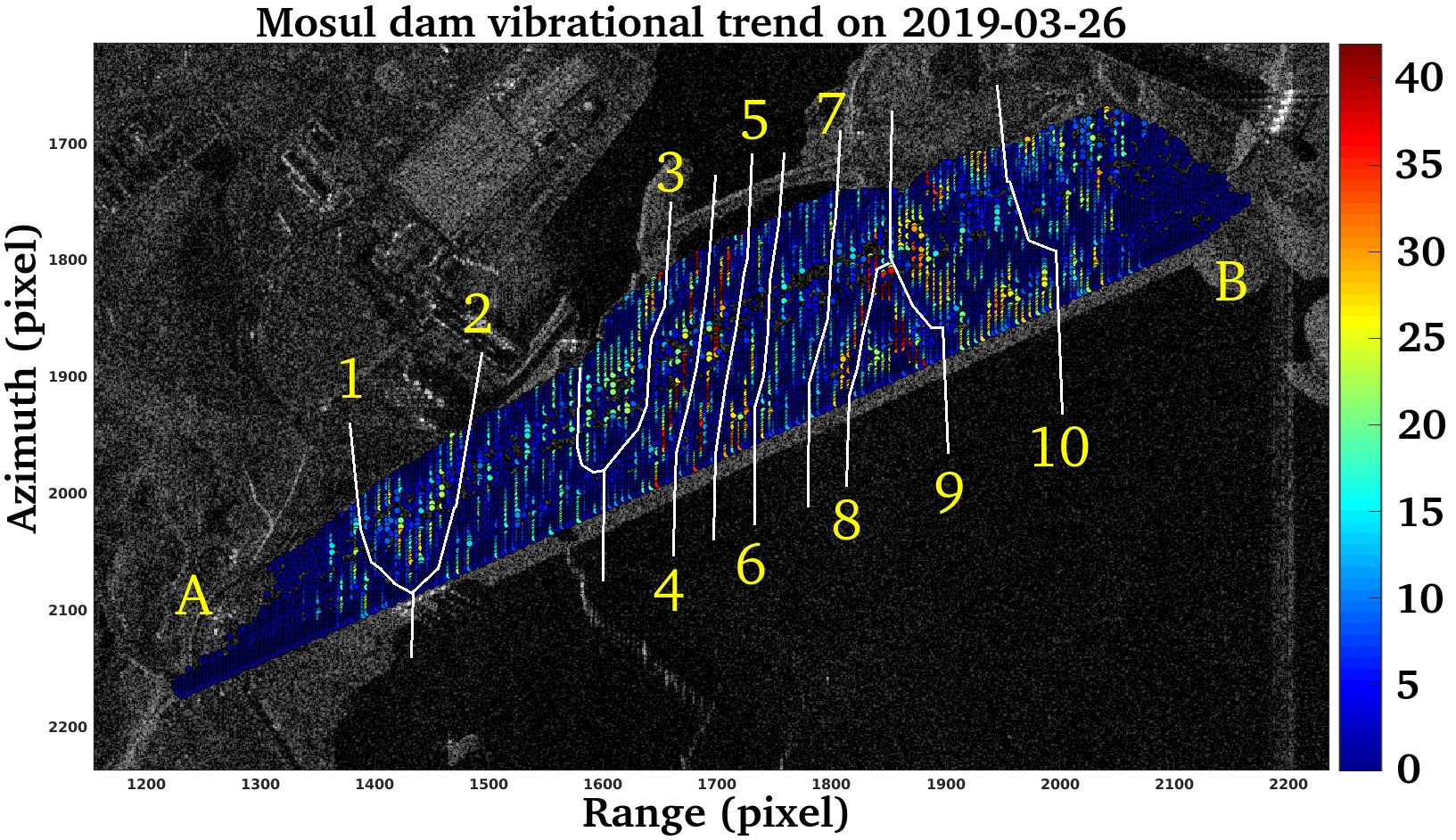}
	\caption{Vibrations estimated on the 26 March 2019 (case study 2-c). Numbers indicate the retrieved cracks. Letters represent the dam's edges.}
	\label{Vibrazioni_4}
\end{figure}

\begin{figure}[htb!]
	\centering
	\includegraphics[width=8.5cm,height=6.5cm]{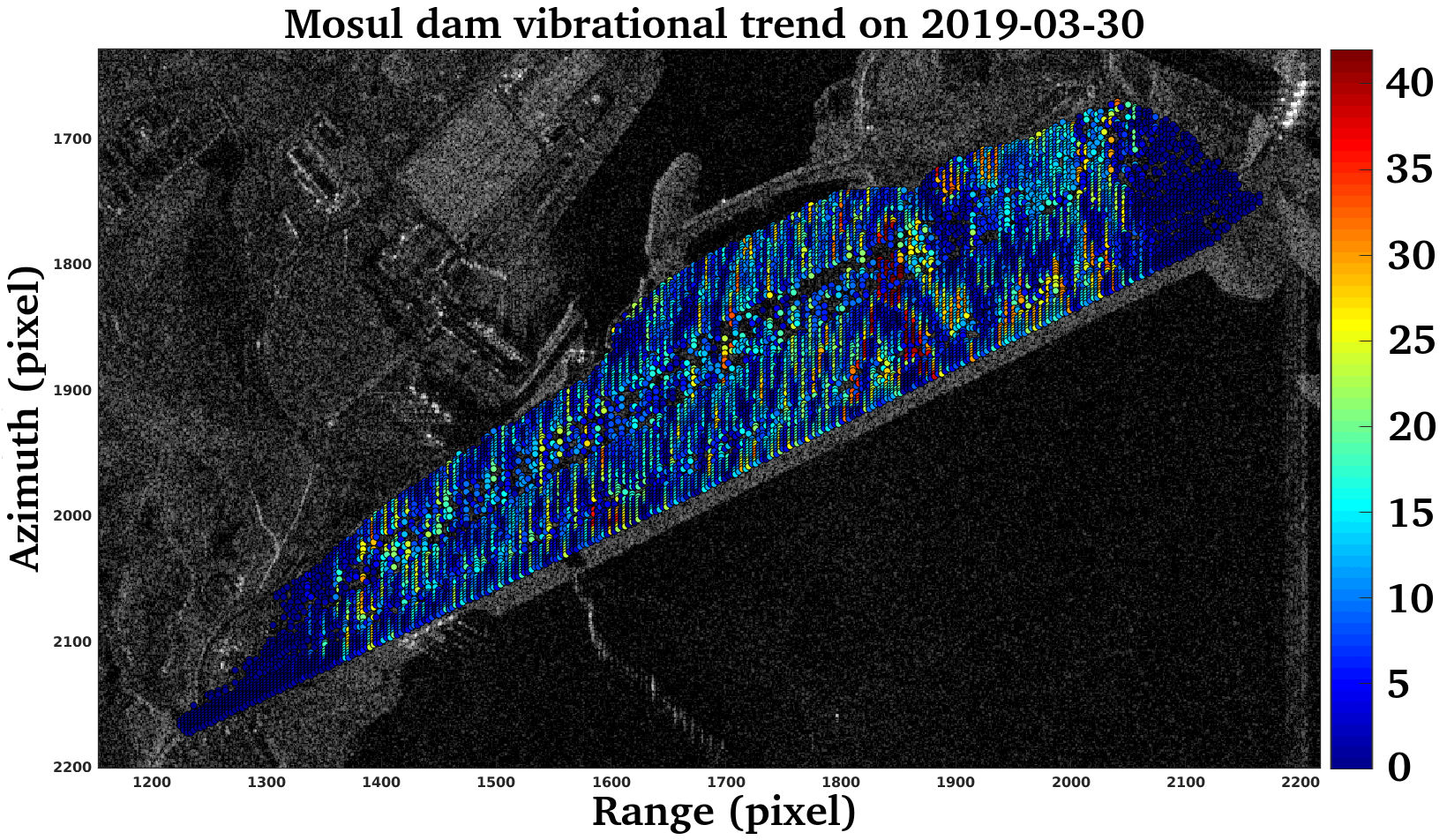}
	\caption{Vibrations estimated on the 30 March 2019 (case study 2-d).}
	\label{Vibrazioni_5}
\end{figure}

\begin{figure}[htb!]
	\centering
	\includegraphics[width=8.5cm,height=6.5cm]{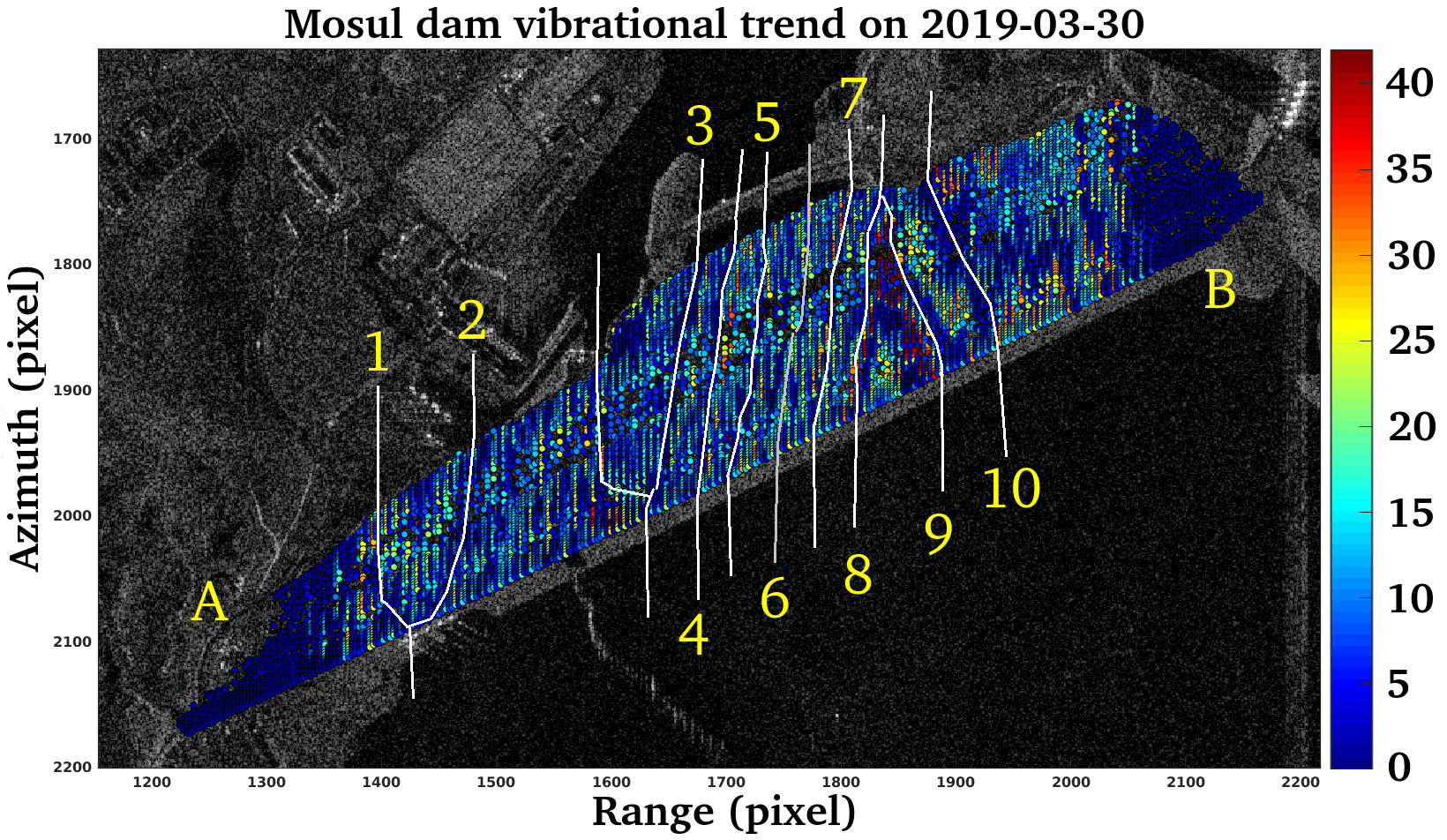}
	\caption{Vibrations estimated on the 30 March 2019 (case study 2-d). Numbers indicate the retrieved cracks. Letters represent the dam's edges. }
	\label{Vibrazioni_6}
\end{figure}

\begin{figure}[htb!]
	\centering
	\includegraphics[width=8.0cm,height=6.0cm]{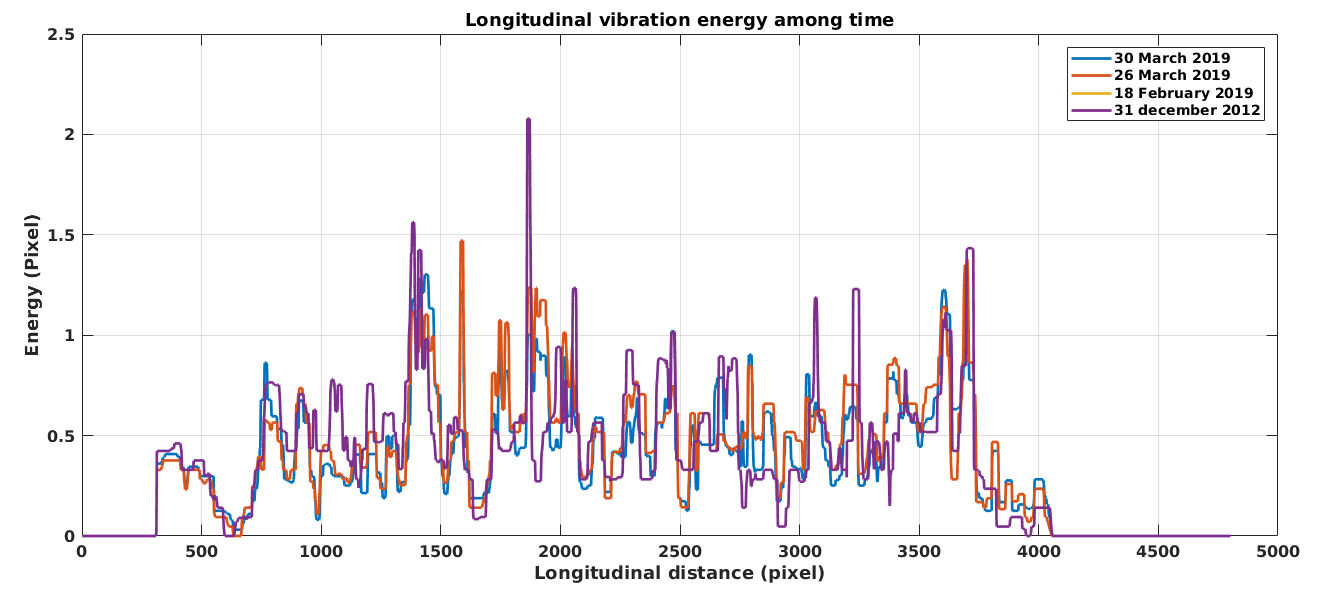}
	\caption{Vibrations estimated on the 30 March 2019 (case study 2-d).}
	\label{Vibrazioni_7}
\end{figure}

\subsection{First case study}\label{Case_Study_one}

In Figure \ref{Bridge_1_1}, the results obtained using the PS-InSAR technique are shown (case study 1-a of Table \ref{Tab_1}). It is possible to notice that a single anomaly, labelled as "2", is found in this case, even if a large number of images is used to produce the analysis.

Figure \ref{Bridge_1} contains the SAR image of the first case study 1-b (see Table \ref{Tab_1}). Measurement points are chosen based on the amplitude dispersion index. Precisely, all the pixels with high energy in magnitude are selected and the displacements are estimated on these points.

From the Modal analysis, two areas are found with vibrational anomalies. These anomalies were detected because they had a much higher vibrational energy with respect to the other candidate pixels. These areas are indicated by the yellow arrows in Figure \ref{Bridge_1}. 

The corresponding optical image is shown in Figure \ref{Bridge_2} where only the anomaly, labelled as "2" is visible, and is characterized by a potential crack of the bridge that has been temporarily repaired by an iron addition, placed on the road surface (this is clear in subfigure (b) that is a zoomed version around the "2" arrow). In the area underlined by the "1" arrow, there is probably a transversal crack, not big enough to be visible with the optical image. This anomaly could be proved after an inspection on the site or through other dedicated in-situ measurements, with the objective to find damages on the infrastructure.

\begin{table*}[tb!]

\caption{The main characteristics of the satellite datasets.}\label{Tab_1}
\begin{tabular}{ p{1.2cm} p{1.9cm} p{1.6cm} p{5.25cm} p{4.3cm} p{1.4cm}}
\hline 
\hline 
\textrm{Case Study} &\textrm{Location} & \textrm{Processing} & COORDINATES (WGS-84) & Time of obs. & Number of obs.\\
\hline
1-a & Mosul bridge & PS-InSAR & 44$^\circ$ 53' 18.84"  N, 11$^\circ$  36' 28.89" E & December 2012 - March 2019 & 83\\
1-b & Mosul bridge & Modal & 44$^\circ$ 53' 18.84"  N, 11$^\circ$  36' 28.89" E & 31 December 2012 & 1\\
2-a & Mosul dam& PS-InSAR & 43$^\circ$ 00' 37.11"	N, 12$^\circ$ 25' 45.15" E & December 2012 - March 2019 & 83\\
2-b & Mosul dam & Modal & 43$^\circ$ 00' 37.11"	N, 12$^\circ$ 25' 45.15" E & 31 December 2012 & 1\\
2-c & Mosul dam& Modal & 43$^\circ$ 00' 37.11"	N, 12$^\circ$ 25' 45.15" E & 26 March 2019 & 1\\
2-d & Mosul dam& Modal & 43$^\circ$ 00' 37.11"	N, 12$^\circ$ 25' 45.15" E & 03 March 2019 & 1\\

\hline
\hline 
\end{tabular}

\end{table*}

\subsection{Second case study}\label{Case_Study_two}

The second case study is aimed at the detection of possible cracks throughout the Mosul 
dam drawing a full understanding of the infrastructure stability situation identifying 
potentially dangerous areas. This case study is divided into two parts, the first part concerns the application of the PS-InSAR technique (namely, experiment 2-a in Table \ref{Tab_1}), while, the second part involves 
the application of the modal analysis (namely, experiments 2-b, 2-c and 2-d in Table \ref{Tab_1}). 

Experiment 2-a consists of displacement measurements performed through the processing of a long time series  of interferometric SAR data. Specifically, the maps concerning displacement, velocities, and acceleration have been estimated and the first order acceleration variations are calculated. Figure \ref{Cumulative_Displacement_all} refers to the displacement map, while Figure \ref{Cumulative_Velocities} is the cumulative velocity map calculated as the 
first derivative approximation. 
The cumulative accelerations are depicted in Figure \ref{Cumulative_Accellerations_All}. Finally, 
Figure \ref{Cumulative_Accellerations_Var_1} shows the acceleration variation. For this figure we stopped 
at the first order of variation. In Figure \ref{Displacement_1}, three displacement trends are represented, 
measured on the $\{1,2,3\}$ points represented in Figure \ref{Cumulative_Displacement_all}. The points cover 
the entire dam in the longitudinal direction. Point $1$ is located on the west side, point $2$ on the center 
and point number 3 on the opposite end, located at east of the dam. From the inspection at 
Figure \ref{Displacement_1}, it can be seen that the maximum difference in height in terms of 
cumulative displacement is about 42 millimeters, compared to the LOS of the radar. The east-west ends of 
the infrastructure are upset while the central part is in subsidence. This also shows that the dam is bending 
like an arch through the long temporal series. Figure \ref{Line_Displacement} shows how much displacement 
is taking place on the L1 line drawn in Figure \ref{Cumulative_Displacement_all}. Figure \ref{Line_Velocity} 
shows the first-order derivative of the displacement, representing the velocity of the persistent scatterers 
as a function of time. This data is always measured on the L1 line. Finally, Figure \ref{Line_Acceleration} 
represents the accelerations, again on L1, as the time displacement second derivative trend. 

All Figures \ref{Vibrazioni_1}-\ref{Vibrazioni_7} represent the estimated vibration 
modes obtained by means of the proposed method applied to a single radar image 
from SLC configuration. To this end, four SAR images, with acquisition dates 
indicated in Table \ref{Tab_1}, have been processed. The first image has been chosen very far from the present day acquired in 2012. 
The others are all dated 2019.  
Looking at all the figures it is possible to see that the estimated vibrational energy concentrates differently 
on the whole dam. 
The dam is vibrating due to the wind and to the pressure of 
the waves generated by the water of the containment basin. 
Under these conditions, the infrastructure oscillations are functions 
of space and time, according to the modes that are located on the 
same areas of the dam, but with different amplitudes, depending on both the intensity of the wind, 
and the current carried by the water of the basin. 
Looking at all oscillation modes, it is reasonable to suppose the presence of cracks, 
small and invisible to remote sensing devices, that extend both longitudinally and transversely over 
the dam \cite{pekau,OWOLABI}. More precisely, the western part of the dam has the identified crack number 1, of pure transversal type, and the recognized crack number 2, is mixed, both longitudinal and transversal. The central part, the most vulnerable, could have 5 cracks, all transversal, numbered as $\{3,4,5,6,7\}$ (see Figures \ref{Cumulative_Accellerations_Var_1}, \ref{Vibrazioni_2}, \ref{Vibrazioni_4} and \ref{Vibrazioni_6}). The eastern part of the infrastructure could include a compound crack, marked with the numbers $\{8,9\}$ and finally, the number 10 appears to be of a purely transversal nature. At the end, the $\{A,B\}$ edges, because they are firmly anchored to the ground, do not generate any vibration.

\section{Discussion}\label{Discussion}
In this section, we elaborate on the experimental results observed in the two case studies and by applying 
PS-InSAR investigation technique as well as the proposed modal analysis also in comparison
with GNSS in-situ measurements contained in \cite{Othman_1}. 
As stated above, the PS-InSAR analysis is useful to highlight accelerations' inversion 
because abrupt variations in the acceleration values could be linked to 
potential cracks in the infrastructure. 
Moreover, the vibrational anomalies estimated through the Modal analysis can 
represent potential cracks in the infrastructures as shown in \cite{pekau,OWOLABI}.

For the first case study, it is possible to state that, even if with a single image, the Modal Analysis is able 
to highlight that the infrastructure oscillates on two points, symmetrically positioned with respect to the center. 
On the western side of the bridge (marked by point 2), there is a narrowing of the roadway consisting of an iron 
scaffold, simply resting on the bridge base. 
This generates a vibrational mode with more energy with respect to the rest of the bridge. 
The symmetrical side, positioned to the east of the Tigris, appears intact, as can be seen from the 
Figure \ref{Bridge_2}. 
This region, indicated by the number $1$ arrow shown in Figure \ref{Bridge_1}, oscillates with 
a fairly high energy. Here, we are quite confident that it is an anomalous oscillation generated 
by the junction of two decks of the bridge. In this case, results from the PS-InSAR are not able to give this information, but it is possible to highlight only the anomaly labeled as number ``$2$''.

\begin{table*}[]
\caption{Comparison with in-situ measurements: the mean velocity of the PS points and from the in-situ GNSS (mm$\cdot$yr$^{-1}$) at the Mosul Dam. Coordinates of the in-situ GNSS and PS and the corresponding distances in [m] are also reported.}
\centering
\begin{tabular}{c|c|c|c|c|c|c}
Lat (GNSS) & Lon (GNSS) & Lat (PS) & Lon (PS) & Distance {[}m{]} & Velocity (GNSS) & Velocity (PS) \\ \hline \hline
36.631111       & 42.815833        & 36.631122     & 42.815912      & 10.04            & -1.99           & -1.82         \\
36.631389       & 42.814722        & 36.631140     & 42.814803      & 07.21            & -0.81           & -0.34         \\
36.631389       & 42.814722        & 36.631140     & 42.814803      & 04.26            & -4.17           & -4.26         \\
36.630278       & 42.818222        & 36.630273     & 42.818258      & 03.00            & -4.17           & -5.02         \\
36.629722       & 42.819444        & 36.629795     & 42.819583      & 16,42            & -4.38           & -4.02         \\
36.629444       & 42.820833        & 36.629513     & 42.820645      & 19,35            & -4.91           & -5.75         \\
36.630278       & 42.821111        & 36.630387     & 42.820935      & 17.63            & -5.13           & -4.83         \\
36.630278       & 42.823889        & 36.628816     & 42.824167      & 65.63            & -5.85           & -5.12         \\
36.629444       & 42.824722        & 36.628843     & 42.825119      & 55.75            & -5.45           & -4.98         \\
36.628333       & 42.824444        & 36.628747     & 42.824168      & 57.61            & -4.8            & -4.45        
\end{tabular}
\label{tabINSITU}
\end{table*}

With focus on the second case study, it is possible to observe that the Mosul dam is still moving and, 
as shown in Figure \ref{Cumulative_Displacement_all}, the maximum concentration of displacement 
is in the center of it; actually, we measure its maximum value at about 4.3 mm. 
The velocity map, reported in Figure \ref{Cumulative_Velocities}, confirms this result: 
the highest estimated speed values are concentrated in the center of the dam and are on 
average equal to 4 mm per year. The analysis of the accelerations reported 
in Figure \ref{Cumulative_Accellerations_All} shows that the dam is divided into two equal parts: 
the L2 symmetry line divides the dam into the western part that is negatively accelerating, 
while the eastern part is doing so, but, positively. 
This phenomenon is also confirmed in the identification of the potential cracks 
number $\{3,4,5,6,7\}$ visible in Figures \ref{Vibrazioni_2}, \ref{Vibrazioni_4} and \ref{Vibrazioni_6}. 
From Figure \ref{Cumulative_Accellerations_Var_1}, it is possible to observe that cracks number $\{1,2\}$ are visible separately, cracks $\{3,4,5,6,7\}$ are contained within the red area located to the east of the infrastructure, finally the cracks number $\{8,9,10\}$ are faithfully detected separately. 
As can be seen from Figure \ref{Line_Displacement}, the estimated displacement is increasing over time and tends to increase more on the central part, slightly shifted to the east side of the dam. 
From Figure \ref{Line_Velocity}, it is possible to detect some anomalies during the period between 2018 and 2019, with respect to the normal trend that goes from 2012 and 2018. According to this figure we realize that the dam is sinking deeper into the central part of it. 
Figure \ref{Vibrazioni_6} is also very important because it restores the position of the cracks previously estimated using the m-m technique, which coincides with the change of sign of the accelerations as a function of time. It is therefore possible to find the cracks number $\{1,3,7,10\}$. 
The oscillation modes in Figure \ref{Vibrazioni_7} are surprisingly the same, even when observed at very long distances. This campaign of measures concerns oscillation modes measured on the 18$^{th}$ February and the 26$^{th}$ and the 30$^{th}$ March 2019, compared to the oscillation modes of the 31$^{th}$ December 2012.  The position of the oscillation modes remains almost unchanged, slightly changing the amplitude.

Finally, to further validate the proposed method, we resort to the GNSS in-situ 
measurements extracted from \cite{Othman_1}.
In Table \ref{tabINSITU}, the displacement velocities measured by both differential GNSS and PS points 
estimates are reported. Particularly, the mean velocity (mm$\cdot$yr$^{-1}$) of the PS points and the mean velocity measured from the in-situ GNSS (mm$\cdot$yr$^{-1}$) are given together with the coordinates and the related distances. 
The PS points are selected in the proximity of the in-situ measurement stations with distances ranging from 3 m to 57.61 m. The estimated velocities show a good agreement with the in-situ measurements resulting in a very high correlation coefficient.

\section{Conclusions}\label{Conclusions}
In this paper, a method exploiting micro-motion estimation is proposed with the specific goal to identify potential cracks existing on the Mosul dam. This method is based on the modal analysis of the extended and elastic bodies typical of large infrastructures. The motivation for this approach resides in the fact that  vibrating modes reflect predictable models for a given structure and radar is a particularly sensitive instrument to vibrations. Results are also compared with those by the PS-InSAR technique. A long temporal series of remote sensed data from the CSK SAR constellation has been processed. Results show clearly the possibility of visualizing the oscillation modes of the Mosul dam, measured from each SLC image, at different points of its main body. Moreover, the vibrations over the dam using the pixel tracking technique have been also estimated.

The results of this study show that, with regard to the analysis based on the PS-InSAR technique, the subsidence displacement of the dam continues to increase, as does the speed. As far as acceleration is concerned, a sudden variation with sign reversal on two longitudinal stretches of the dam has been estimated. This phenomenon suggests the presence of several cracks occurred on the Mosul dam. With regard to the results obtained through the modal analysis, several cracks are displayed at least coinciding with those estimated using the PS-InSAR technique. Thus, the results of the proposed analysis highlight that the modal analysis is a viable tool to estimate the m-m of critical infrastructures. This application is further confirmed by the fact that it is not necessary to acquire a long time series, but only one image in the SLC can be sufficient to obtain reliable results.

Recent research efforts also highlight the robustness of the proposed methodology against the SAR observation geometry, Particularly, experiments with observations varying from a ''right-ascending'' to a ''right-descending'' configuration, do not entail any particular distortion and/or limitation to the visibility of these singularities on infrastructures.

\section{ACKNOWLEDGMENTS}
This research used the software SARPROZ: https://www.sarproz.com/, for performing the multi-temporal atmospheric phase screen estimation. The authors would like to thank the Italian Space Agency (ASI) for providing the SAR data.

\balance
\bibliographystyle{IEEEtran}
\bibliography{group_bib}
\end{document}